\documentclass[twocolumn, times]{aastex631}

\usepackage{booktabs}
\usepackage{multirow}
\usepackage{amsmath}
\hypersetup{colorlinks=true, citecolor=blue, urlcolor=blue, linkcolor=blue, allcolors = blue}

\begin{document}

\title{COSMOS2025: Machine Learning Classification of Early- and Late-type Galaxies at $0 < \text{z} < 3$}


\author{Vahid Asadi~\href{https://orcid.org/0009-0005-8897-2385}{\includegraphics[scale=0.04]{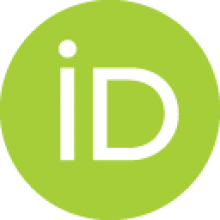}}}
\affiliation{Department of Physics, Institute for Advanced Studies in Basic Sciences (IASBS), PO Box 11365-9161, Zanjan, Iran; \url{vahijd.asadij@gmail.com}}

\author{Najmeh Sheikhi~\href{https://orcid.org/0009-0009-4165-6376}{\includegraphics[scale=0.04]{orcid-ID.png}}}
\affiliation{Department of Physics, Institute for Advanced Studies in Basic Sciences (IASBS), PO Box 11365-9161, Zanjan, Iran; \url{vahijd.asadij@gmail.com}}


\begin{abstract}
We present a fast, interpretable machine learning framework to classify early- and late-type galaxies in the COSMOS2025 catalog at $0<\text{z}<3$, without relying on image-based training labels or computationally expensive structural fitting. Using the Santa Cruz Semi-Analytic Model, we generate a training set with secure morphological labels defined by bulge-to-total mass ratio and specific star formation rate. We bridge the simulation-to-observation domain gap by injecting realistic photometric noise derived from COSMOS2025. A \texttt{CatBoostClassifier} trained on 66 broadband colors achieves excellent performance in the simulated domain, recovering late-types with $98\%$ precision/recall and early-types with $91\%$ precision and $88\%$ recall. Applied to 44,132 COSMOS2025 galaxies, the model reveals a striking bimodality: only $\sim$6\% of galaxies receive intermediate probabilities ($0.3 < \text{P}(\mathrm{\text{Early type}}) < 0.7$)—nearly identical to the fraction observed in the simulation. This demonstrates that broadband colors are a decisive morphological discriminant, with the remaining 94\% classified at high confidence. Validation against independent bulge+disk decompositions yields $70\%$ overall accuracy, with late-types identified at $78\%$ purity and $74\%$ completeness. The most important color feature, F277W--F444W, reflects the expected optical/NIR contrast between old and young stellar populations. The full pipeline completes in under 30 minutes on standard hardware, demonstrating that simulation-trained color-based classifiers offer a scalable, physically interpretable route to approximate morphology for large next-generation surveys.
\end{abstract}



\section{Introduction}
The morphological classification of galaxies is fundamental to our understanding of galaxy formation and evolution. Since the inception of the Hubble sequence \citep{hubble1926extragalactic}, the structural dichotomy between early-type galaxies--typically quiescent, bulge-dominated spheroids--and late-type galaxies--typically star-forming, disk-dominated systems--has served as a primary probe of the physical processes driving cosmic history. This morphological bimodality correlates with other physical properties, such as color, stellar mass, and star formation rate \citep[e.g., ][]{strateva2001color, bell2003optical, baldry2004quantifying}, suggesting that morphology traces the integrated merger history and feedback mechanisms acting upon a galaxy \citep[e.g., ][]{conselice2014evolution, somerville2015physical}.

While morphological classification is well-established in the local universe ($\text{z} \sim 0$), extending these studies to higher redshifts ($\text{z} > 1$) presents significant challenges. At these epochs, galaxies are intrinsically smaller, fainter, and often exhibit irregular structures that defy classical Hubble types \citep[e.g., ][]{mortlock2013redshift, huertas2016mass}. Traditional methods, such as visual classification \citep[e.g., ][]{lintott2008galaxy, kartaltepe2015candels, simmons2016galaxy, willett2016galaxy} or parametric fitting of surface brightness profiles \citep[e.g., ][]{peng2002detailed, simard2011catalog,vika2013megamorph,dimauro2018catalog,nedkova2024bulge+,shuntov2025cosmos2025}, become increasingly computationally expensive and subject to biases caused by surface brightness dimming and point spread function effects in large-volume surveys. Furthermore, the advent of next-generation surveys with the James Webb Space Telescope \citep[JWST; ][]{gardner2006james}, Euclid \citep[e.g., ][]{cuillandre2025euclid}, and the Nancy Grace Roman Space Telescope \citep[e.g., ][]{wang2022high} will yield datasets of unprecedented volume, rendering manual or distinct computationally intensive fitting techniques impractical for the entirety of the observed populations.

To address these data challenges, machine learning (ML) has emerged as a practical tool for astronomical classification. In recent years, ML algorithms---most notably convolutional neural networks (CNNs)---have been applied to separate galaxy populations morphologically based on imaging data \citep[e.g.,][]{dominguez2018improving,cheng2021galaxy,bhambra2022explaining,fang2023automatic,iyer2024galaxy,Cao2024galaxy,luo2025galaxy,pandya2025sidda}. While powerful, CNN-based methods face several intrinsic limitations for large-scale, multi-survey applications at high redshift. Their performance is heavily dependent on the specific imaging data (e.g., depth, resolution, point spread function, and bandpasses) on which they are trained, making them difficult to generalize consistently across different surveys or even within a single survey as observational conditions vary \citep[see discussions in, e.g., ][]{barchi2020machine}. Furthermore, the feature representations learned by CNNs are often opaque ``black boxes", providing limited physical insight into the photometric drivers of morphological classification \citep[e.g., ][]{csahin2025unlocking}. Most critically, the majority of these methods are supervised learning approaches that require large, consistently labeled training sets, which at high redshifts ($\text{z} > 1$) are scarce, potentially biased by visual classification at low signal-to-noise, and may not reflect the true physical diversity of galaxies during peak assembly epochs.  

In this study, we present an approach that leverages photometric colors rather than pixel-level images, trained on physically grounded simulations rather than limited observational labels. We construct a robust ML framework to classify early- and late-type galaxies out to $\text{z} \sim 3$ using the COSMOS2025 catalog \citep{shuntov2025cosmos2025}. Our method utilizes the updated Santa Cruz Semi-Analytic Model (SAM) \citep{somerville2021mock, yung2022semi} to generate a large, high-redshift training set with intrinsic morphology labels defined by bulge-to-total mass ratio (B/T) and specific star formation rate (sSFR). Rather than assigning rigid binary labels, our classifier outputs a continuous probability that quantifies a galaxy's photometric similarity to the early-type archetype. This probabilistic framework naturally accommodates galaxies with intermediate or mixed properties.

We treat the ``simulation-to-observation" domain gap by injecting realistic photometric noise derived from the COSMOS2025 data itself. By training a \texttt{CatBoostClassifier} model \citep{prokhorenkova2018catboost} on photometric colors—features, we obtain a classifier that is computationally efficient, and whose decisions can be traced to physically meaningful color indices.

The paper is organized as follows. In Section \ref{sec:2}, we describe the mock and observational datasets. Section \ref{sec:3} details the pre-processing steps, including missing value imputation and noise injection. Section \ref{sec:4} outlines the ML architecture and training strategy. We present the performance metrics and results in Sections \ref{sec:5} and \ref{sec:6}, followed by a discussion of the implications and limitations in Section \ref{sec:7}. We conclude in Section \ref{sec:8}.

We adopt a flat $\Lambda$CDM cosmology with parameters $\text{H}_{0}=70\text{kms}^{-1}\text{Mpc}^{-1}$, $\Omega_{\text{m}}=0.3$ and $\Omega_{\Lambda}=0.7$. All magnitudes are reported in the AB system \citep{oke1983secondary}.


\section{Data}\label{sec:2}
This study employs two complementary datasets: SAM mock galaxy data, which provides a physically motivated training sample with known intrinsic galaxy properties, and the COSMOS2025 observational catalog \citep{shuntov2025cosmos2025}, which serves as the target dataset for early- and late-type galaxy classification. In this section, we describe both datasets and the selection criteria applied to construct the working samples.

\subsection{Mock Data}\label{sec:2.1}
For this work we use JWST wide-field light-cone catalogs constructed from the updated Santa Cruz SAM galaxy formation model \citep{somerville2021mock,yung2022semi}. The model evolves galaxies within dark matter merger trees extracted from the Bolshoi--Planck $N$-body simulation \citep{klypin2011dark}, assuming a $\Lambda$CDM cosmology. Baryonic processes are treated with parameterized prescriptions for gas accretion and cooling, star formation, stellar and supernova feedback, black hole growth and AGN feedback, and galaxy mergers, yielding self-consistent predictions for key galaxy properties such as stellar mass, star formation rate, metallicity, and SEDs.

To mimic deep extragalactic surveys, the SAM outputs are organized into light cones that trace the evolving galaxy population across cosmic time. The catalogs cover five independent fields with footprints matched to the CANDELS legacy fields \citep{grogin2011candels,koekemoer2011candels}: GOODS-S, GOODS-N, COSMOS, EGS, and UDS. Each field contains multiple realizations spanning $0 \leq \text{z} \leq 10$, and includes observed-frame photometry for a wide range of facilities (e.g. JWST/NIRCam, Roman/WFI, HST/WFC3 and ACS, Spitzer, Euclid, Rubin, GALEX, SDSS, UKIRT, VISTA, and DECam), as well as rest-frame luminosities in UV and optical bands \citep{yung2019semi,somerville2021mock,yung2022semi}. Physical quantities for halos and galaxies, including stellar and bulge masses, are provided and form the basis for our morphology definition and training labels.

In the SAM, galaxies are assigned to dark matter halos as centrals and satellites. Gas accretion, cooling, and star formation follow an updated Kennicutt–Schmidt relation tied to the molecular gas content, with an additional starburst mode associated with galaxy mergers \citep{robertson2006fundamental,hopkins2009disks,somerville2015star}. Stellar and AGN feedback can eject or heat gas, regulating subsequent star formation and black hole growth \citep{somerville2008semi,bondi1952spherically}. Model parameters are calibrated to reproduce observed galaxy mass and luminosity functions and key scaling relations, and the resulting SEDs are processed through dust attenuation and IGM absorption models before being convolved with instrument response functions to produce realistic broadband fluxes and magnitudes \citep{madau1995radiative,yung2019semi,somerville2021mock,yung2022semi}.

\subsection{Observation Data}\label{sec:2.2}
The observational component of this work is based on the COSMOS2025 galaxy catalog \citep{shuntov2025cosmos2025}, the definitive data release from the COSMOS-Web JWST Treasury program (GO\#1727; PIs: Casey \& Kartaltepe). COSMOS2025 builds on previous COSMOS compilations \citep{laigle2016cosmos2015,weaver2022cosmos2020} and provides photometry, photometric redshifts, morphological measurements, and derived physical parameters for over 700{,}000 galaxies in the central $\sim 0.54\,\mathrm{deg}^{2}$ of the COSMOS field \citep{scoville2007cosmic}.

The catalog combines deep, high-resolution imaging from both space- and ground-based facilities. JWST/NIRCam (F115W, F150W, F277W, F444W) and JWST/MIRI (F770W) data from COSMOS-Web are complemented by HST/ACS F814W imaging, together with extensive ground-based ultraviolet (CFHT/MegaCam), optical (Subaru/HSC and Suprime-Cam), and near-infrared data from UltraVISTA DR6 (VISTA/VIRCAM; \citealt{mccracken2012ultravista}). Source detection is performed on a PSF-homogenized $\chi^{2}$ combination of the four NIRCam bands, and photometry is measured in a homogeneous way across 37 bands covering $0.3$–$8\,\mu$m.

Morphological and photometric measurements are obtained using \textsc{SExtractor++}, which fits Sérsic and bulge+disk models to the native-resolution images convolved with the appropriate PSF, delivering consistent total fluxes and structural parameters across all bands. Photometric redshifts and physical parameters are derived with \texttt{LePhare} \citep{arnouts1999measuring,ilbert2006accurate} and \texttt{CIGALE} \citep{boquien2019cigale}, using 32 bands spanning $0.3$–$8\,\mu$m. The combination of deep JWST/NIRCam imaging, expanded template libraries with diverse star formation histories and dust attenuation curves, and improved ground-based data from UltraVISTA DR6 and HSC PDR3 yields photometric redshift accuracies of $\sigma_{\mathrm{NMAD}} \sim 0.01$–$0.03$ down to faint magnitudes and significantly improved stellar-mass completeness compared to COSMOS2020 \citep{weaver2022cosmos2020}. 

In this work, we use the \textsc{SExtractor++} model-fit photometry, together with the associated \texttt{LePhare} photometric redshifts and stellar masses, to construct our observational sample.

\subsection{Sample Selection}\label{sec:2.3}
For a fair comparison between the SAM and COSMOS2025 catalogs, we first identified a common set of broad-band filters available in both datasets. We selected twelve bands spanning the ultraviolet to near-infrared, which provide strong leverage on galaxy colors, stellar masses, and star formation histories. These include the CFHT u band, HST/ACS F814W, UltraVISTA $\text{YJHK}_{\text{s}}$, JWST/NIRCam F115W, F150W, F277W, F444W, and \textit{Spitzer}/IRAC channels 1 and 2, as summarized in Table~\ref{tab:tab1}.

\begin{table}[t]
\centering
\caption{Overview of the twelve COSMOS2025 bands used in this work.}
\setlength{\belowcaptionskip}{10pt}
\begin{tabular}{ccccc}
	\hline \hline
	Instrument                & Band  & Central\footnote{Median of the transmission curve.} & Width\footnote{Full width of the transmission curve at half maximum.} & Depth\footnote{5$\sigma$ depth in empty apertures with diameters of $1.0^{\prime\prime}$ for ground-based, $0.15^{\prime\prime}$ for JWST/NIRCam and HST/ACS, and $0.5^{\prime\prime}$ for JWST/MIRI images, averaged over the NIRCam area.} \\
	/Telescope                &       & $\lambda$ [{\AA}] & [{\AA}] &      \\
	(Survey)                  &       &                   &         &      \\
	\hline
	CFHT                      & u         &  3858           &  598    & 27.3 \\
	\hline
	NIRCam                    & F115W       & 11622           & 2646    & 27.2 \\
	& F150W       & 15106           & 3348    & 27.4 \\
	& F277W       & 28001           & 6999    & 28.1 \\
	& F444W       & 44366           & 11109   & 28.0 \\
	\hline
	ACS/HST                   & F814W       & 8333            & 2511    & 27.5 \\
	\hline
	VIRCAM                    & Y         & 10216           & 923     & 25.8 \\
	/VISTA                    & J         & 12525           & 1718    & 25.8 \\
	UltraVISTA                & H         & 16466           & 2905    & 25.5 \\
	DR6                       & $\text{K}_{\text{s}}$     & 21557           & 3074    & 25.3 \\
	\hline
	IRAC                      & ch1         & 35686           & 7443    & 26.4 \\
	/\textit{Spitzer}         & ch2         & 45067           & 10119   & 26.3 \\
	\hline
	\label{tab:tab1}
\end{tabular}
\end{table}

To construct our working samples, we then applied identical cuts to both the mock and observational catalogs:

\begin{itemize}
	\item[--] Redshift range: $0 < \text{z} < 3$. This range is well sampled in both catalogs and covers the epoch where our B/T-based morphology classification is most reliable.
	
	\item[--] Stellar mass cut: $\log(\text{M}_{\star}/\mathrm{\text{M}}_{\odot}) > 9.5$. This threshold avoids the dwarf regime, where our B/T morphology indicator becomes less reliable.
	
	\item[--] Magnitude cut: $m_{\mathrm{\text{F444W}}} < 27.5$. This limit is set by the depth and photometric quality of COSMOS2025 catalog.
\end{itemize}

This selection results in a final sample of 265{,}504 galaxies from the SAM catalog (restricted to the COSMOS light cone and using all seven available realizations) and 44{,}132 galaxies from the COSMOS2025 catalog.


\section{Pre-processing}\label{sec:3}
To enable a consistent machine-learning analysis between the idealized mock catalog and the real observational data, the selected samples from Section~\ref{sec:2} must be harmonized and transformed. This section details the steps taken to address missing data, construct informative color features, inject realistic observational noise into the simulations, and define robust morphology labels for model training.

\subsection{Filling Photometry Missing Values}\label{sec:3:1}
The SAM sample contained no missing values in the selected bands (Table~\ref{tab:tab1}). In contrast, the COSMOS2025 sample exhibited missing data in all bands except F444W, which was used for sample selection (Table~\ref{tab:tab11}).

\begin{table}[t]
	\centering
	\caption{Percentage of missing values for magnitudes and fluxes in the COSMOS2025 sample across the twelve selected bands.}
	\setlength{\belowcaptionskip}{10pt}
	
	\begin{tabular}{c|c|c}
		\hline
		\hline
		COSMOS2025 selected band & \multicolumn{2}{c}{Missing value [\%]} \\
		\cmidrule{2-3}
		
		& Magnitude & Flux \\
		\hline
		u        & 7.50     & 0.00 \\
		\hline
		F115W    & 0.45     & 0.00 \\
		F150W    & 0.08     & 0.00 \\
		F277W    & 0.01     & 0.00 \\
		F444W    & 0.00     & 0.00 \\
		\hline
		F814W    & 0.88     & 0.00 \\
		\hline
		Y        & 0.96     & 0.00 \\
		J        & 0.39     & 0.00 \\
		H        & 0.15     & 0.00 \\
		$\text{K}_{\text{s}}$  & 0.09     & 0.00 \\
		\hline
		ch1      & 0.23     & 0.00 \\
		ch2      & 0.25     & 0.00 \\
		
		\hline
	\end{tabular}
	\label{tab:tab11}
\end{table}

ML-based imputation methods have become increasingly prevalent in astronomy for handling missing photometric data, offering advantages over traditional interpolation or mean-substitution approaches by capturing complex, non-linear relationships in multi-band datasets \citep[e.g.,][]{stekhoven2015missforest,ren2020using, keerin2022estimation, luo2024imputation}. Among these methods, \texttt{MissForest} \citep{stekhoven2015missforest} is particularly well-suited for astronomical applications because it is a non-parametric, iterative ensemble method that makes no assumptions about the underlying data distribution. It naturally handles high-dimensional interactions between bands and is robust to outliers, allowing for the reconstruction of missing entries without requiring extensive feature engineering or architectural tuning.

To address the missing data in this study, we employed the \texttt{MissForest} imputation method, an iterative approach based on the Random Forest algorithm \citep{Breiman2001}. In this procedure, each band containing missing values is modeled as a function of the other bands. In each iteration, a Random Forest is trained on the observed data to predict the missing entries. This process cycles repeatedly through the affected bands, with imputed values from one step informing the predictions in the next, until the solution converges (i.e., changes become minimal between iterations).

This approach captures the complex, non-linear relationships between photometric bands, ensuring the imputed magnitudes are statistically consistent with the underlying multi-wavelength structure of the complete dataset.

To assess the impact of imputation on our results, we performed a sensitivity analysis using the complete SAM sample as ground truth. We first transferred the observed missingness pattern from the COSMOS2025 catalog to the SAM sample in a magnitude-dependent manner. We divided the COSMOS2025 galaxies into magnitude bins based on their F444W brightness---a band that is effectively 100\% complete. Within each bin, we calculated the fraction of galaxies missing measurements in every other photometric band.

These calculated, bin-specific missing fractions were then applied as probabilities to the complete SAM sample: galaxies in a given magnitude bin had measurements in each band removed randomly according to the corresponding observational missing fraction. This method ensured that the artificially introduced missingness in the SAM sample realistically reflected the survey's detection limits, where fainter galaxies are more likely to have missing measurements---particularly in less sensitive bands---rather than imposing a uniform missing fraction.

After imputing the incomplete SAM dataset using \texttt{MissForest}, we evaluated the imputation accuracy by comparing the imputed values to the true SAM values, calculating the mean absolute error (MAE) for each band. Results demonstrate that imputation is highly accurate (Table~\ref{tab:imputation_accuracy}): MAE ranges from 0.026 mag for the u-band (8.9\% missing) to essentially zero ($< 10^{-5}$ mag) for NIR bands (F277W, F444W, F150W, all with $<0.1\%$ missing).

\begin{table}
	\centering
	\caption{Missing fraction and imputation accuracy (MAE) for each band in the SAM sensitivity test.}
	\setlength{\belowcaptionskip}{10pt}
	\begin{tabular}{lcc}
		\hline \hline
		Band & Missing value [\%] & MAE \\
		\hline
		u & 8.90 & $3 \times 10^{-2}$ \\
		\hline
		F115W & 0.70 & $3 \times 10^{-4}$ \\
		F150W & 0.10 & $3 \times 10^{-5}$ \\
		F277W & 0.00 & $8 \times 10^{-7}$ \\
		F444W & 0.00 & 0.0 \\
		\hline
		F814W & 1.10 & $1 \times 10^{-3}$ \\
		\hline
		Y & 1.70 & $8 \times 10^{-4}$ \\
		J & 0.80 & $8 \times 10^{-4}$ \\
		H & 0.30 & $6 \times 10^{-5}$ \\
		$\text{K}_{\text{s}}$ & 0.20 & $6 \times 10^{-5}$ \\
		\hline
		ch1 & 0.50 & $1 \times 10^{-4}$ \\
		ch2 & 0.60 & $4 \times 10^{-5}$ \\
		
		\hline
	\end{tabular}
	\label{tab:imputation_accuracy}
\end{table}

\begin{figure*}
	\centering
	\includegraphics[width=0.95\linewidth]{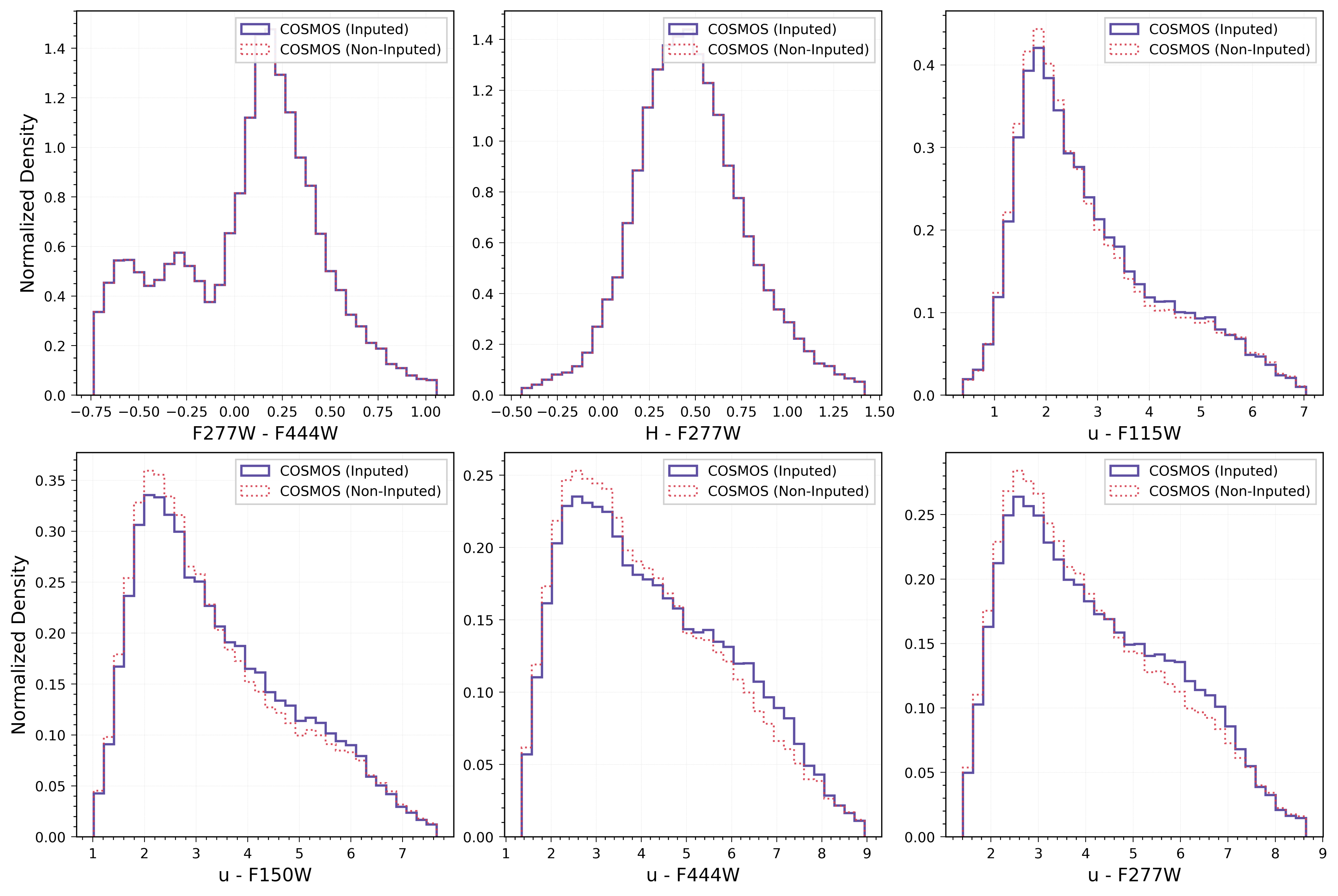}
	\caption{Comparison of color-index distributions for galaxies in the COSMOS2025 sample with imputed photometry (solid blue histograms) and with original (non-imputed) photometry (dotted red histograms).
	}
	\label{fig:color_distributions_imputed}
\end{figure*}

Additionally, Figure~\ref{fig:color_distributions_imputed} compares the distributions of the six most important color indices (see Figure~\ref{fig:fig8}) for the COSMOS2025 sample galaxies with imputed versus non-imputed photometry. The histograms nearly overlap, confirming that imputation does not bias the feature space. This verifies that the imputation process preserves the underlying photometric relationships and introduces no meaningful bias to the feature space used for classification.

\subsection{Constructing Colors}\label{sec:3.2}
After filling missing values of the selected bands, we derived colors from the twelve photometric bands to serve as the primary features for our ML model. To encapsulate the SED shape comprehensively and enable the model to capture complex, non-linear relationships, we generated all unique pairwise color combinations. This yielded a total of 66 distinct colors, encompassing both short wavelength-baseline combinations (e.g., F227W - F444W, J - H) and long baseline combinations (e.g., u - F115W, $\text{K}_\text{s} - \text{ch1}$).

We chose to use color indices rather than raw apparent magnitudes for some key reasons. Colors provide a more direct and robust diagnostic of galaxy physical properties. While apparent magnitudes are sensitive to distance and flux calibration uncertainties, color indices are intrinsically distance-independent and mitigate many such systematic effects. Furthermore, because colors measure the relative flux between bands, they more effectively constrain the SED shape---information that is critical for differentiating between stellar populations, star formation histories, and dust attenuation characteristics \citep[see for more detail][]{luo2024photometric}.

Each color probes a distinct physical regimen: UV--optical colors diagnose young stellar populations and ionized gas, optical--NIR colors reveal stellar mass and dust properties, and long-baseline colors encode integrated SED information. By including all 66 pairwise combinations, we ensure that no potentially informative diagnostic is omitted, while allowing our ML model to learn which combinations are most relevant for morphological classification.

\begin{figure*}
	\centering
	\includegraphics[width=0.95\linewidth]{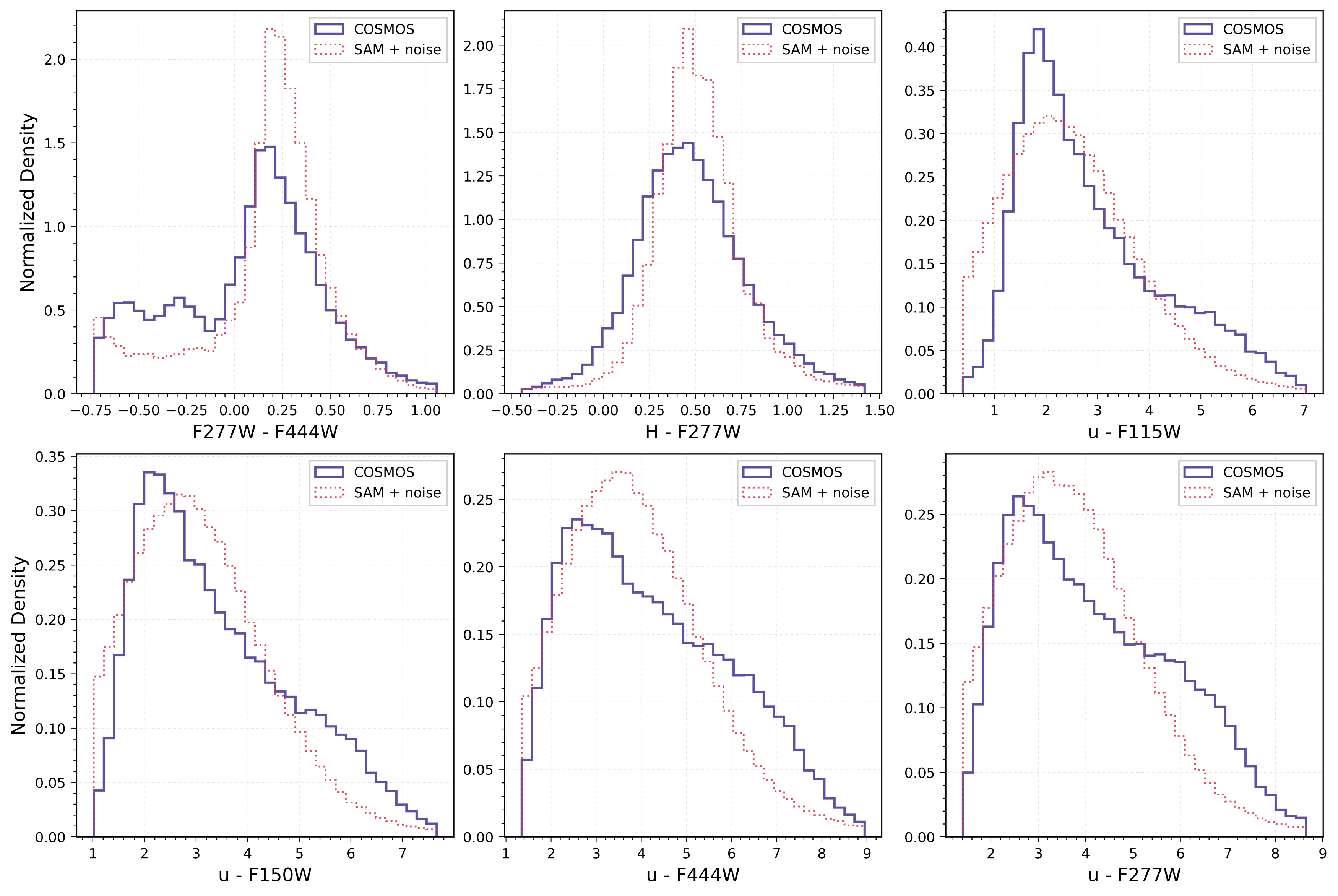}
	\caption{
		Comparison of color distributions between the COSMOS2025 sample (blue solid histograms) and the noise-injected SAM sample (red dotted histograms).
	}
	\label{fig:hist}
\end{figure*}

\subsection{Injecting Observational Noise}\label{sec:3:3}
To ensure that the idealized photometry from the SAM sample is realistically compatible with the observational COSMOS2025 sample, we injected noise to mimic its observed color-error properties. This was implemented in a three-step process:

\begin{itemize}
	\item[1.] For each of the 66 colors in the COSMOS2025 sample, we trained a \texttt{RandomForestRegressor} model \citep{Breiman2001} to predict the color error using the color value itself as the sole feature.
	
	\item[2.] We applied these 66 trained models to the corresponding colors in the pristine SAM sample, generating a predicted color error ($\sigma_{C,\text{pred}}$) for every object and color.
	
	\item[3.] To simulate observational uncertainties, we added a random perturbation, $\delta C$, to each SAM color. Each $\delta C$ was drawn from a Gaussian distribution with a mean of zero and a standard deviation equal to the predicted error for that specific object and color \citep{asadi_mass}.
\end{itemize}

In Figure~\ref{fig:hist}, the normalized histograms of several important representative colors (see Figure~\ref{fig:fig8}) illustrate that the SAM+noise sample (dotted curves) reproduces both the overall shapes and dynamic ranges of the COSMOS2025 distributions (solid curves), with only modest residual shifts in peak positions and tails.

\subsection{Setting Labels}\label{sec:3.4}
We construct high-purity morphological labels for the SAM sample using a two-stage approach that combines structural (B/T ratio) and star-formation-activity indicators. Traditional morphological classifications often rely on a single bulge-to-total mass ratio threshold (typically around \(\mathrm{\text{B/T}} \simeq 0.4\)) to separate early- and late-types \citep[e.g., ][]{de1996near,allen2006millennium}; however, such a purely structural cut can mix quenched disks and star-forming bulges in the intermediate regime, reducing label purity for our ML method. Instead, we adopt a more conservative strategy that requires consistency between galaxy structure and star-formation state, explicitly trading completeness for cleaner, physically homogeneous training labels.

\begin{figure*}
	\centering
	\includegraphics[width=\linewidth]{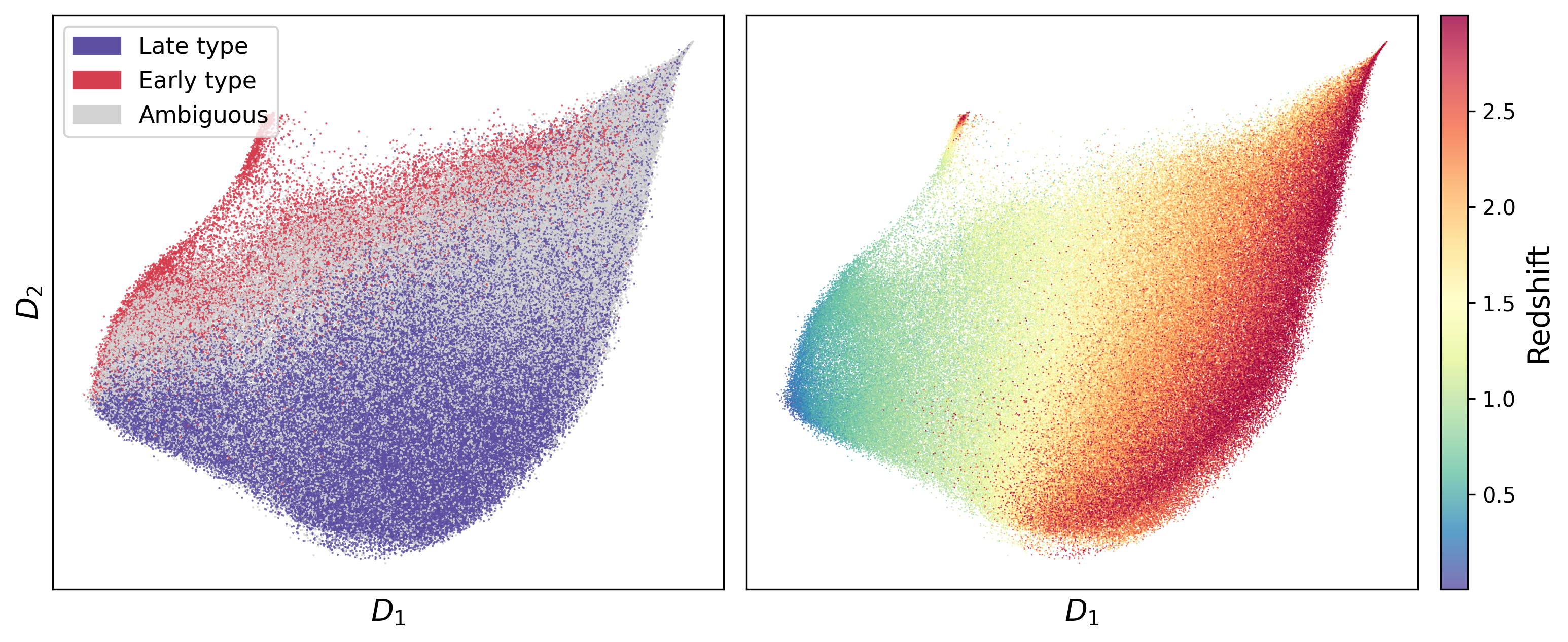}
	\caption{UMAP projection of the SAM sample in color space. Left: Classification labels showing secure late-type (blue), secure early-type (red), and ambiguous (grey) galaxies. Right: The same projection colored by redshift.}
	\label{fig:fig1}
\end{figure*}

First, to identify the quiescent and star-forming populations at different epochs, we adopted the evolving sSFR threshold from \cite{pacifici2016evolution}, defined as:

\begin{equation}
	\text{sSFR} \leq \text{sSFR}_{lim} = \frac{0.2}{\text{t}_{U}(\text{z})}
	\label{eq1}
\end{equation}

Here, sSFR is in \(\text{Gyr}^{-1}\), and \(\text{t}_{U}(\text{z})\) is the age of the universe at redshift z in Gyr. This criterion provides a dynamic threshold that evolves with cosmic time, reflecting the observed decline in galaxy star-formation activity.

Second, we classify galaxies based on their bulge-to-total stellar mass ratio:
\begin{equation}
	\mathrm{\text{B/T}} = \frac{\text{M}_{\text{bulge}}}{\text{M}_{\star}},
\end{equation}
using strict thresholds:
\begin{itemize}
	\item [--] Early-type candidate: $\mathrm{\text{B/T}} \geq 0.5$ (bulge-dominated)
	\item [--] Late-type-type type candidate: $\mathrm{\text{B/T}} \leq 0.3$ (disk-dominated)
\end{itemize}
Galaxies with zero or non-finite $M_{\text{bulge}}$ are treated as pure disks ($\mathrm{\text{B/T}} = 0$).

We then combine these indicators to define secure labels:
\begin{itemize}
	\item [--] Secure early-type: $\mathrm{\text{B/T}} \geq 0.5$ and classified as quiescent
	\item [--] Secure late-type: $\mathrm{\text{B/T}} \leq 0.3$ and classified as star-forming
	\item [--] Ambiguous: All other galaxies (intermediate B/T or morphology–sSFR mismatch)
\end{itemize}

This yields 105{,}600 secure late-type (41\%), 20{,}391 secure early-type (8\%), and 139{,}513 ambiguous galaxies (51\%). Only the secure populations are used for ML training to ensure high label purity.

Our decision to train exclusively on secure early- and late-type galaxies is a deliberate methodological choice. By anchoring the ML classifier on these two physically distinct extremes, we define a ``morphological–star-formation axis'' in color space. The classifier's primary output---when applied to the full SAM or COSMOS2025 samples---is the continuous probability $\text{P}(\mathrm{\text{Early type}})$, which quantifies a galaxy's photometric similarity to the secure early-type archetype relative to the secure late-type archetype. This strategy serves two primary scientific purposes:

\begin{itemize}
	\item[--] As a validation probe: it tests whether the photometric distinction learned from the secure extremes extrapolates meaningfully into the transitional regime. Does structural ambiguity imply photometric ambiguity?
	\item[--] As a continuum diagnostic: the classifier assigns $\text{P}(\mathrm{\text{Early type}})$ to every galaxy (in both the SAM and COSMOS2025 samples), placing each object along a continuous early–late axis defined by the secure archetypes, rather than forcing a binary classification (the hard threshold at $\text{P}(\mathrm{\text{Early type}}) = 0.5$ is presented only for compatibility).
\end{itemize}

Figure~\ref{fig:fig1} shows a 2D Uniform Manifold Approximation and Projection \citep[UMAP; ][]{mcinnes2018umap} of the SAM sample galaxies built from the 66 colors: the left panel is colored by our morphology labels (secure late-type, secure early-type, and ambiguous), and the right panel shows the same projection colored by redshift. Secure late-types and secure early-types occupy broadly distinct regions of the manifold, while ambiguous systems predominantly fill the interface between them, tracing a continuous transition in color space. The redshift-colored panel reveals a smooth gradient across the manifold, with low-redshift galaxies concentrated toward one side and higher-redshift galaxies toward the other, demonstrating that the color-based embedding encodes both morphological and redshift information.

\section{Machine Learning Method}\label{sec:4}
The core of our analysis is to first train a ML classifier on the 66 noise-injected colors of the secure early- and late-type subsample, reproducing the labels defined in Section~\ref{sec:3.4}, and then use this model to obtain both probabilistic and hard classifications for the full SAM sample (including galaxies flagged as ambiguous) and for COSMOS2025.

\subsection{CatBoost Classifier}\label{sec:4.1}
We employed the \texttt{CatBoostClassifier} algorithm \citep{prokhorenkova2018catboost} to perform supervised classification of galaxies into early- and late-type classes. CatBoost has demonstrated appropriate performance in astronomical applications, including galaxy classification \citep{Asadi_2025}, quasar identification \citep{hughes2022quasar}, and source characterization \citep{coronado2022classification}, among other studies \citep{humphrey2023euclid, cunha2022photometric, coronado2023redshift, zeraatgari2024exploring, boulet2024catalogue, li2025application}.

As a gradient-boosting method, CatBoost builds an ensemble of decision trees sequentially, with each new tree trained to correct the misclassifications of the previous trees, progressively minimizing a chosen loss function and improving the overall predictive performance of the classifier. Importantly for our 66-dimensional color feature space, CatBoost is inherently robust to feature redundancy: the ensemble mechanism automatically prioritizes informative features and down-weights collinear or uninformative ones. This robustness allows us to include all 66 color combinations without risk of overfitting or loss of generalization---the model naturally filters redundancy and identifies the physically most relevant diagnostics.

We selected CatBoost over deep learning alternatives (e.g., attention-based or recurrent networks) primarily for its interpretability---feature importance rankings directly reveal which colors drive morphological classification---and computational efficiency, while maintaining appropriate performance on our dataset.

\subsection{Splitting the SAM Sample}\label{sec:4.2}
\begin{figure}
	\centering
	\includegraphics[width=\linewidth]{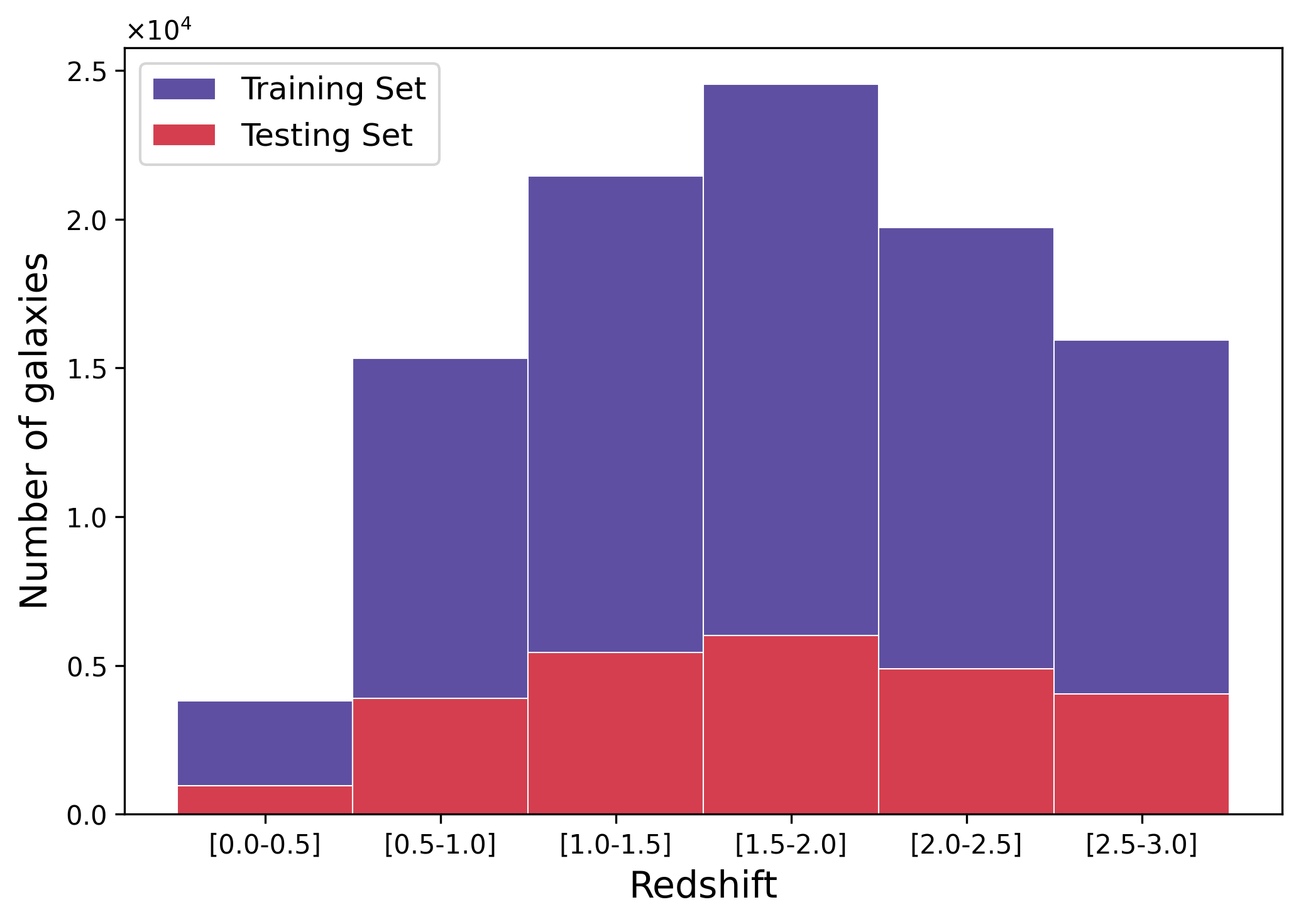}
	\caption{Validation of the training–testing split. The redshift distributions of the training and testing subsets are nearly identical, confirming that the random stratified sampling produces representative and unbiased splits of the SAM sample.}
	\label{fig:fig2}
\end{figure}

\begin{figure*}
	\centering
	\includegraphics[width=\linewidth]{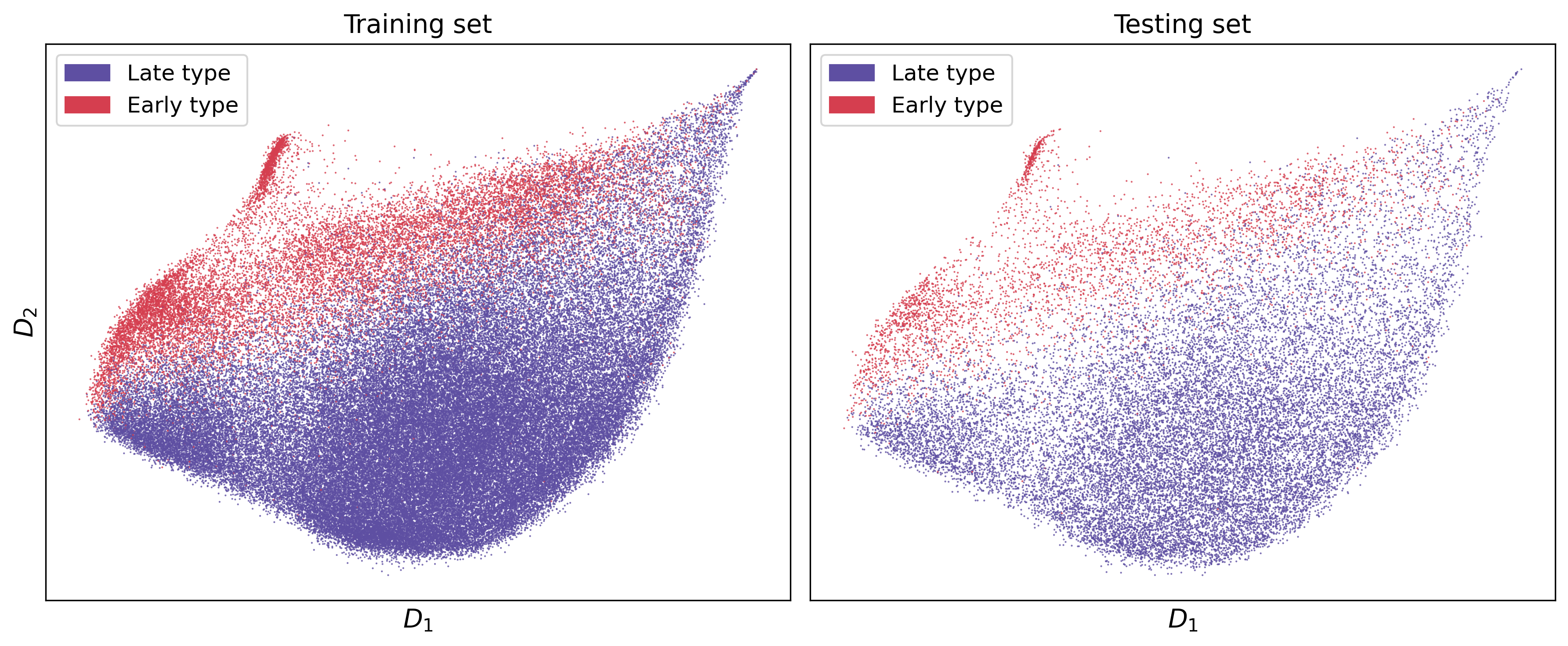}
	\caption{UMAP projection of the secure SAM sample, comparing the training and testing sets obtained from the random stratified split. The overlapping spatial distributions demonstrate that the partition is representative, ensuring that a model trained on the training set can be fairly evaluated on the testing set.}
	\label{fig:fig3}
\end{figure*}

To evaluate the intrinsic performance of the \texttt{CatBoostClassifier} in a controlled setting, we work entirely within the SAM before moving to the observational data. We perform a random, stratified split of the secure SAM sample into a training set (80\%) and a testing set (20\%), with stratification based on the morphology labels to preserve the relative early- and late-type class proportions in both subsets. This yields a training set of 100{,}792 galaxies (including 16{,}313 secure early-types) and a testing set of 25{,}199 galaxies (including 4{,}078 secure early-types). The redshift distributions of the two subsets are nearly identical (Figure~\ref{fig:fig2}), and a 2D UMAP projection (Figure~\ref{fig:fig3}) shows that the training and testing samples occupy the same region of color space, confirming that the split does not introduce obvious selection biases.

\subsection{Optimization}\label{sec:4.3}
The model hyperparameters were tuned using a randomized search with 5-fold stratified cross-validation over 50 iterations \citep{pedregosa2011scikit}, maximizing the F1-score (Equation~\ref{eq:5}). The search focused on three key parameters—the number of iterations, learning rate, and tree depth—which primarily control model complexity, convergence behavior, and the bias–variance trade-off in gradient-boosted decision trees \citep{geron2019hands}. The optimal values obtained from this procedure are listed in Table~\ref{tab:tab2}.
Adopting these optimal hyperparameters, we retrained the \texttt{CatBoostClassifier} on the secure SAM training set and then applied the resulting model to the secure testing set to assess its classification performance.

\begin{table}[t]
	\centering
	\caption{Hyperparameter tuning results: search ranges and final selected values.}
	\setlength{\belowcaptionskip}{10pt}
	
	\begin{tabular}{c|ccc}
		\hline
		\hline
		Hyperparameter & Iterations & Learning rate & Depth \\
		\hline
		Value range & 100--2000 & 0.01--0.3 & 3--10 \\
		\hline
		Optimum value & 932 & 0.05 & 4  \\
		\hline
		
	\end{tabular}
	\label{tab:tab2}
\end{table}

\section{Accuracy Metrics}\label{sec:5}
To evaluate and compare the performance of our ML classifier on the secure SAM testing set, we employ standard metrics derived from the confusion matrix \citep{stehman1997selecting}: precision (purity), recall (completeness), and the F1-score.

Precision measures the accuracy of positive predictions, or the purity of the selected positive class:
\begin{equation}
	\text{Precision} = \frac{\text{TP}}{\text{TP} + \text{FP}}.
	\label{eq:3}
\end{equation}

Recall measures the fraction of all actual positives that are correctly identified, or the completeness of the selection:
\begin{equation}
	\text{Recall} = \frac{\text{TP}}{\text{TP} + \text{FN}}.
	\label{eq:4}
\end{equation}

In the specific context of this work, a positive prediction corresponds to the identification of an early-type galaxy. A high precision therefore indicates a pure sample of early-type galaxies, in which most of the identified candidates are truly early-type, while a high recall indicates a complete sample, meaning the model successfully recovers a large fraction of the true early-type population.

The F1-score, defined as the harmonic mean of precision and recall, provides a single metric that balances the trade-off between purity and completeness:
\begin{equation}
	\text{F1-Score} = 2 \times \frac{\text{Precision} \times \text{Recall}}{\text{Precision} + \text{Recall}}.
	\label{eq:5}
\end{equation}

\section{Result}\label{sec:6}
\subsection{Performance on the SAM sample}\label{sec:6.1}
Having established our classification framework, we now evaluate the performance of the ML classifier within the controlled environment of the secure SAM testing set, where the true morphological labels are known. This allows us to quantify how well the model recovers early- and late-type galaxies before applying it to the observational COSMOS2025 sample.

\begin{figure}
	\centering
	\includegraphics[width=0.9\linewidth]{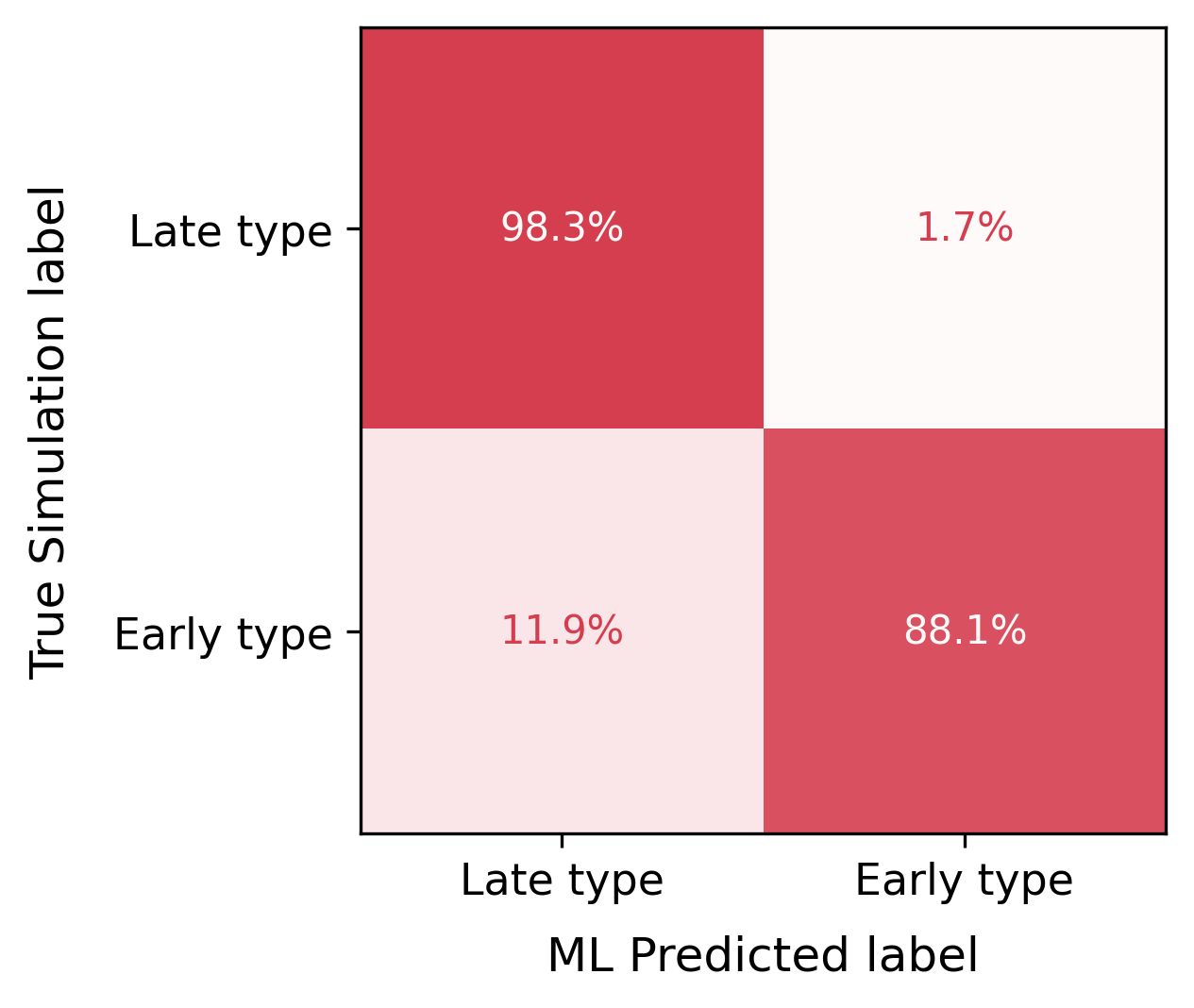}
	\caption{Confusion matrix for the ML classifier evaluated on the secure SAM testing set.}
	\label{fig:fig4}
\end{figure}

The confusion matrix for the classifier, visualized in Figure~\ref{fig:fig4}, provides a clear view of its performance on the secure SAM testing set. For true late-type galaxies, the model correctly recovers 98\% as late-type while misclassifying only 2\% as early-type, indicating an extremely pure late-type prediction channel. For true early-type galaxies, the classifier correctly identifies 88\% as early-type and assigns 12\% to the late-type class, demonstrating strong but slightly less complete recovery of early-type systems compared to late-types.

A more detailed quantitative view is presented in Table~\ref{tab:tab4}, which lists the precision, recall, and F1-score for both morphological classes. For late-type galaxies, the classifier achieves a precision of 98\%, a recall of 98\%, and an F1-score of 98\%, indicating that late-type predictions are both extremely pure and nearly complete. For early-type galaxies, the model attains a precision of 91\%, a recall of 88\%, and an F1-score of 90\%, demonstrating robust performance on the minority class with only a modest reduction in completeness relative to the late-type population.

\begin{table}[t]
	\centering
	\begin{tabular}{c|ccc}
		\hline \hline
		Class & Precision [\%] & Recall [\%] & F1-Score [\%] \\
		\hline
		Late-type  & 97.7 & 98.3 & 98.0 \\
		Early-type & 90.8 & 88.1 & 89.5 \\
		\hline
	\end{tabular}
	\caption{Performance of the ML classifier on the secure SAM testing set for early- and late-type galaxies, evaluated using precision (purity), recall (completeness), and F1-score.}
	\label{tab:tab4}
\end{table}

\begin{figure}
	\centering
	\includegraphics[width=0.9\linewidth]{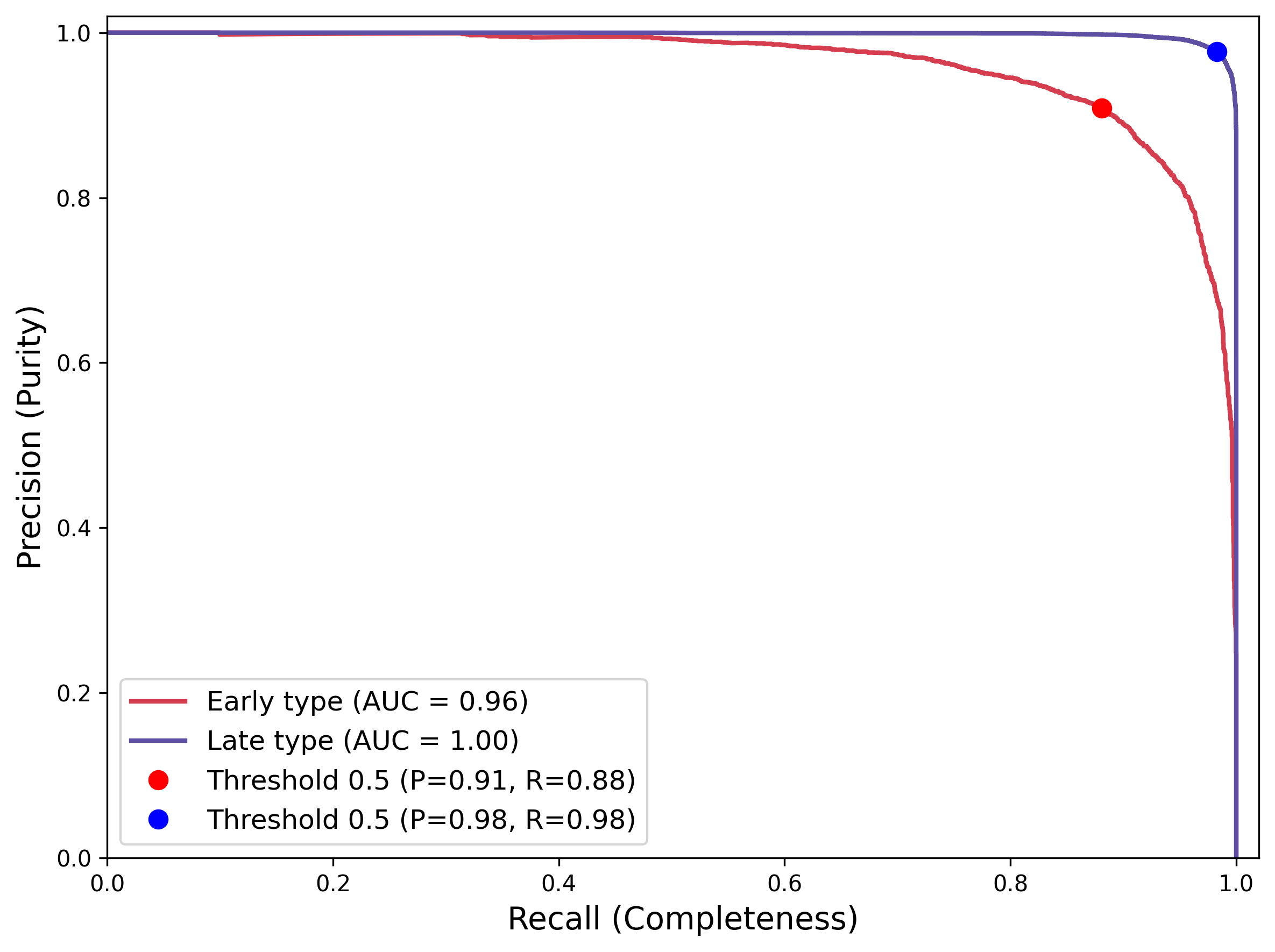}
	\caption{Precision–recall curves for the ML classifier on the secure SAM testing set, shown separately for early-type (red) and late-type (blue) galaxies. Filled circles indicate the default probability threshold of 0.5.}
	\label{fig:fig5}
\end{figure}

The trade-off between purity and completeness in the classifier is further illustrated by the class-specific precision–recall curves shown in Figure~\ref{fig:fig5}. For early-type galaxies, the curve has an Area Under the Curve (AUC) of 0.96, while for late-type galaxies the AUC reaches 1.00, highlighting the model’s excellent overall performance in both regimes. The default probability threshold of 0.5, marked by the colored points in Figure~\ref{fig:fig5}, provides a good balance between precision and recall for each class and corresponds to the summary metrics reported in Table~\ref{tab:tab4}.

\begin{figure*}
	\centering
	\includegraphics[width=\linewidth]{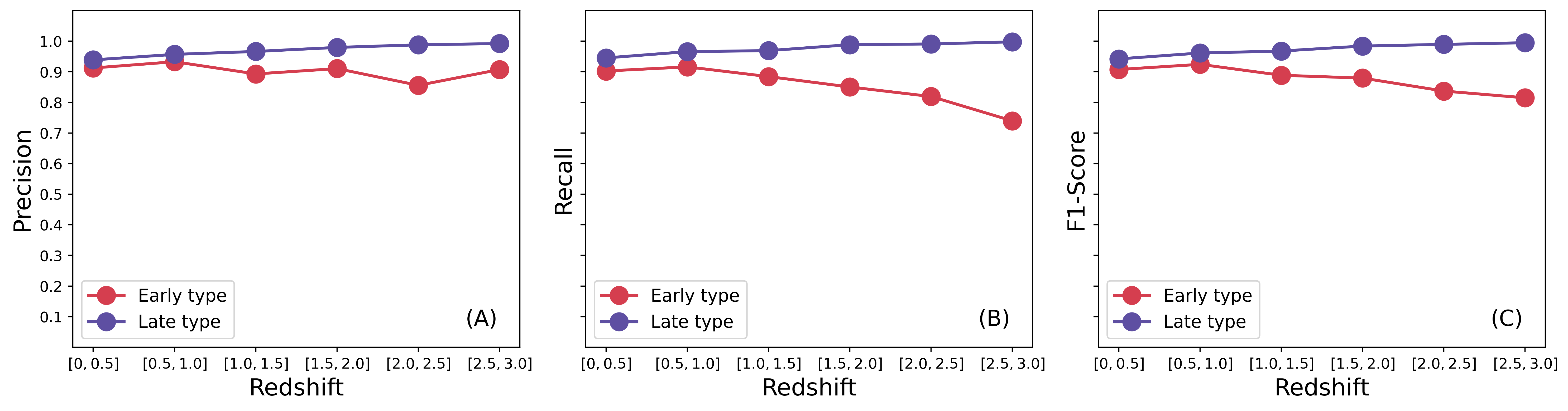}
	\caption{Redshift dependence of the classifier performance on the secure SAM testing set.}
	\label{fig:fig6}
\end{figure*}

While the global metrics demonstrate that the classifier performs very well on average, a more nuanced view emerges when examining performance as a function of redshift. Figure~\ref{fig:fig6} shows the precision, recall, and F1-score for early- and late-type galaxies across six redshift bins in the range $0 < \text{z} < 3$, highlighting how the model behaves as galaxy populations evolve with cosmic time.

\begin{figure}
	\centering	
	\includegraphics[width=\linewidth]{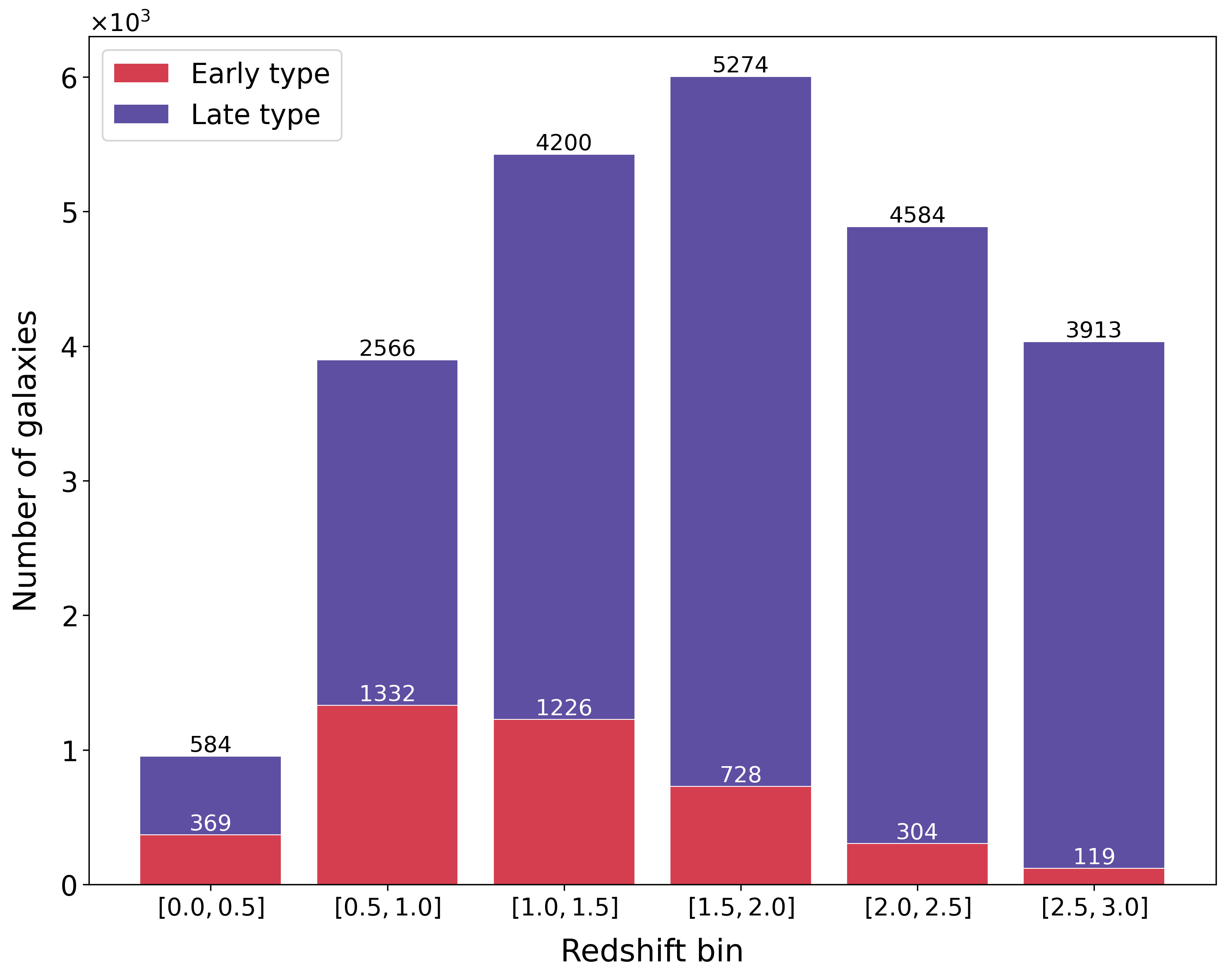}
	\caption{Redshift distribution of secure early- (red) and late-type (blue) galaxies in the SAM test sample, showing the dominance of late types and the declining early-type fraction toward higher redshift.}
	\label{fig:fig55}
\end{figure}

For late-type galaxies, the performance is essentially saturated at all redshifts: precision rises from $\sim 94\%$ to $\sim 99\%$, recall from $\sim 95\%$ to $\sim 100\%$, and F1-score from $\sim 94\%$ to $\sim 99\%$, indicating both extremely pure and nearly complete late-type classifications in every bin. In contrast, the early-type class shows a mild but systematic degradation with redshift: precision remains high ($\sim 91$–93\% at $\text{z}<2$ and $\sim 86$–91\% at $\text{z}>2$), while recall declines from $\sim 90$–92\% in the lowest bins to $\sim 74\%$ in the highest bin, leading to F1-scores that decrease from $\sim 91$–92\% at $\text{z}<1.5$ to $\sim 81\%$ at $2.5 \leq \text{z} < 3.0$. The slight decline in early-type recall at higher redshifts likely reflects the increasing prevalence of irregular and clumpy structures at early cosmic times, where the traditional early-late dichotomy becomes less distinct \citep[e.g., ][]{conselice2014evolution,guo2015clumpy}, combined with the fact that securely early-type systems become intrinsically rare compared to late-types in these bins (see Figure~\ref{fig:fig55}). Nevertheless, the classifier maintains high purity even at these epochs.

\begin{figure*}
	\centering
	\includegraphics[width=0.9\linewidth]{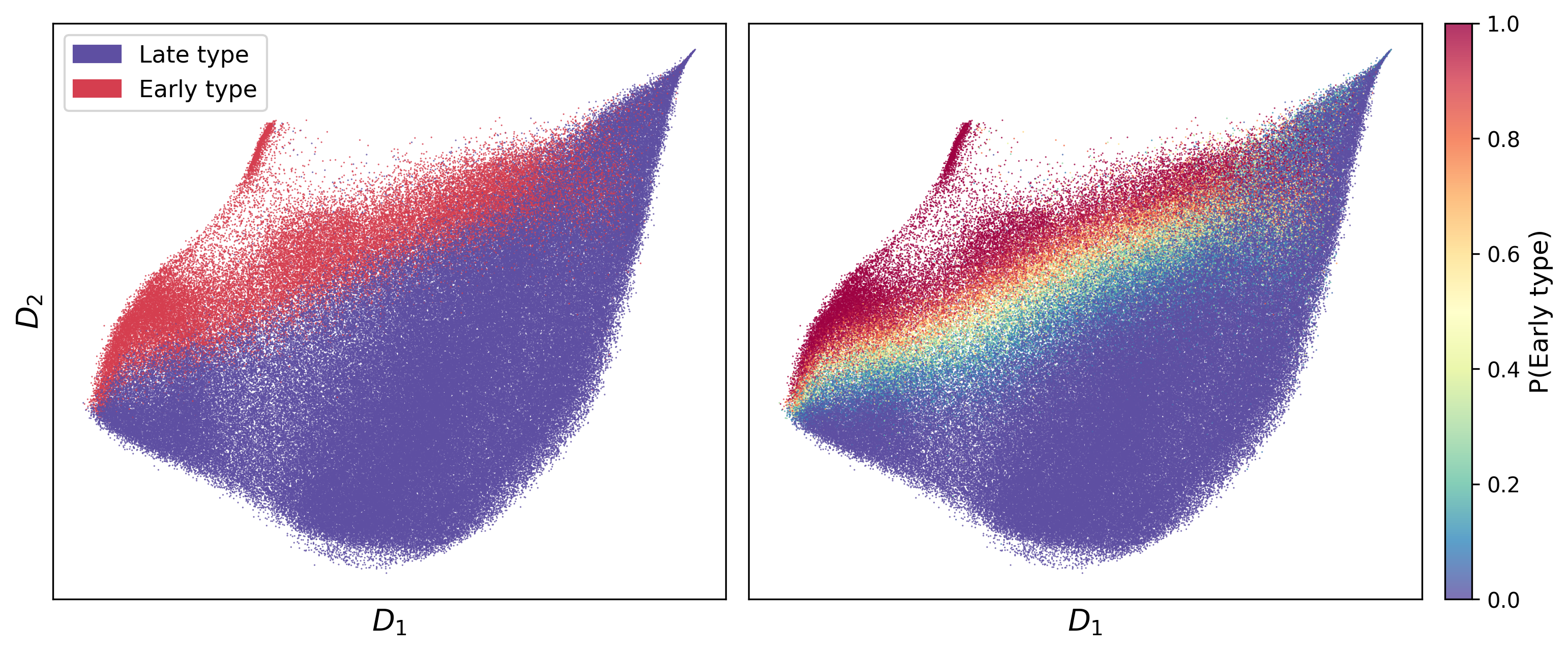}
	\caption{UMAP projection of the full SAM sample in color space after applying the trained \texttt{CatBoostClassifier} on the secure SAM sample. Left: Hard classifications into early-type (red) and late-type (blue) galaxies. Right: The same projection colored by the predicted probability of being early-type.}
	\label{fig:fig7}
\end{figure*}

Having validated the classifier’s performance and confirmed its stability on the held-out testing set, we proceeded to the final classification stage. To ensure the model captured the widest possible range of galaxy properties, we retrained the \texttt{CatBoostClassifier} on the combined training and testing subsets of the secure SAM sample. This final production model, which benefits from the maximum number of available high-confidence labels while maintaining the same hyperparameters and 66-color feature set, was then applied to all SAM galaxies—including those initially flagged as ambiguous. This yields both hard morphology labels (using the default probability threshold of 0.5) and probabilistic classifications in the form of $\text{P}(\mathrm{\text{Early type}})$ for every object, resulting in 222{,}523 galaxies classified as late-type and 42{,}981 as early-type. Figure~\ref{fig:fig7} illustrates the corresponding decision structure in the 2D UMAP space: the left panel shows the hard early- and late-type labels assigned by the classifier, while the right panel displays the early-type probabilities, revealing a smooth gradient across the main color locus that connects clearly late-type and clearly early populations.

Examining the distribution of $\text{P}(\mathrm{\text{Early type}})$ across the full SAM sample reveals a striking bimodality: only 5.8\% of galaxies receive intermediate probabilities ($0.3 < \text{P}(\mathrm{\text{Early type}}) < 0.7$), while the remaining 94.2\% are classified with high confidence ($\text{P}(\mathrm{\text{Early type}}) \leq 0.3$ or $\geq 0.7$). This demonstrates that broadband colors are a decisive morphological discriminant in the simulation domain. As we show in Section~\ref{sec:6.3}, this same bimodality appears in the observational COSMOS2025 sample with nearly identical fractions.

\subsection{Feature Behavior}\label{sec:6.2}
\begin{figure}
	\centering
	\includegraphics[width=\linewidth]{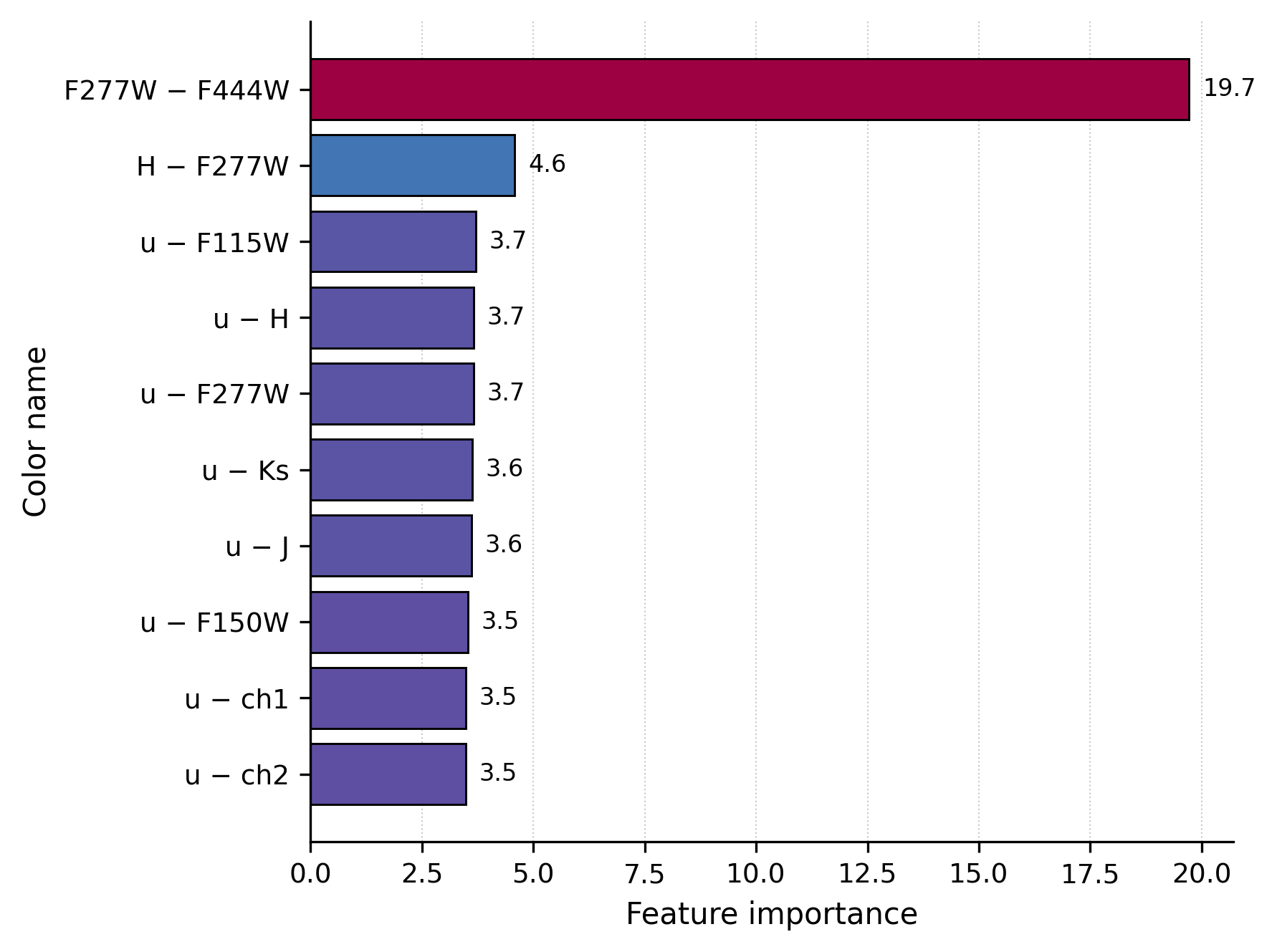}
	\caption{Top ten color features ranked by importance in the \texttt{CatBoostClassifier} trained on the secure SAM sample.}
	\label{fig:fig8}
\end{figure}

Before applying the trained classifier to the COSMOS2025 sample, we first examine how the model uses color information within the SAM, in order to understand which features drive the early- and late-type separation. Figure~\ref{fig:fig8} shows the \texttt{CatBoostClassifier} feature importance ranking for the top ten colors, derived from the final model trained on the secure SAM sample. The most informative feature is the long-baseline NIRCam color F277W-F444W, followed by a set of optical–NIR and UV–NIR colors (e.g. H-F277W, u-F115W, u-H). 

While the full set of 66 colors naturally introduces some redundancy, CatBoost's feature importance mechanism demonstrates that only a subset of features drives the classifier's decisions. The top 10 colors account for 52.6\% of the total gain in the loss function, while the bottom 30 features contribute only 12.1\%, confirming that the model naturally filters uninformative features and is not compromised by the high dimensionality.

CatBoost’s feature importance scores quantify how much each color contributes to reducing the classification loss across the ensemble of trees. In practice, the algorithm aggregates the gain in the objective function each time a feature is used to split a node, summing these contributions over all trees and normalizing them to yield relative importance values \citep{prokhorenkova2018catboost}. This procedure naturally highlights colors that consistently create purer early- and late-type splits.

These feature importance trends have clear physical implications. The dominance of the F277W-F444W color reflects the strong contrast between rest-frame optical and near-IR light, which is sensitive to the relative contribution of old, red stellar populations versus younger, bluer stars; bulge-dominated early-type galaxies are therefore expected to be redder in this color \citep[e.g., ][]{bell2003optical}. The prominence of UV/optical–NIR colors (e.g. H-F277W, u-F115W, u-H, u-F277W, u-$\text{K}_\text{s}$, u-ch1, u-ch2) is likewise consistent with early-types having suppressed recent star formation and stronger 4000\,\AA\ breaks, while late-types show bluer UV-to-NIR gradients due to ongoing star formation \citep[e.g., ][]{poggianti1997indicators,kauffmann2003stellar}.

\begin{figure}
	\centering
	\includegraphics[width=\linewidth]{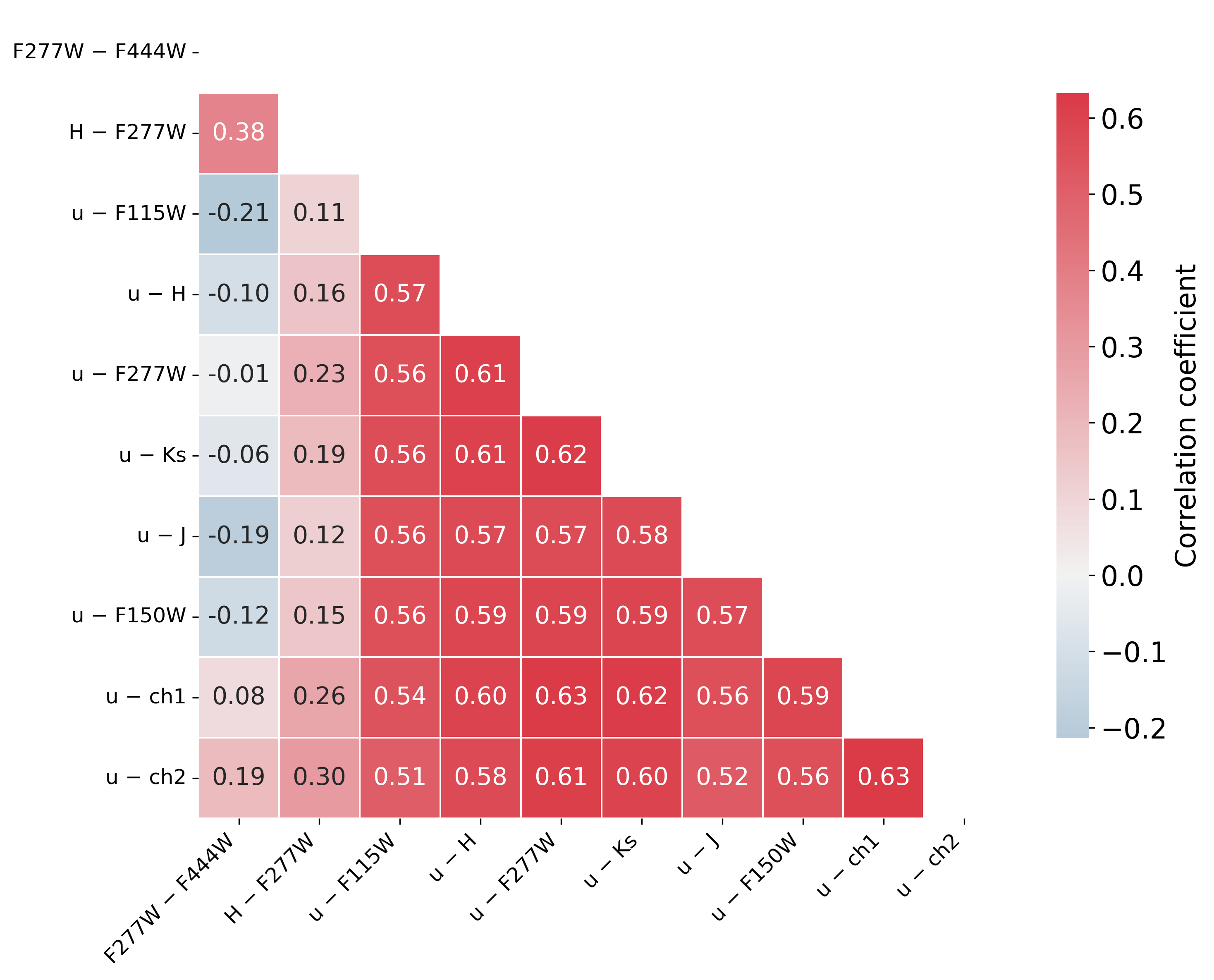}
	\caption{Pearson correlation matrix of the ten most important color features (from Figure~\ref{fig:fig8}), computed from the SAM sample. The diagonal elements (self-correlations) are omitted for visual clarity.}
	\label{fig:corr_matrix}
\end{figure}

While the importance ranking highlights individual colors, understanding the correlations among them provides deeper insight into the underlying physical drivers. To assess this, we computed the Pearson correlation matrix for the top ten features; the result is shown in Figure~\ref{fig:corr_matrix}. The u-based colors form a tightly correlated group with mutual coefficients $r \simeq 0.54$--$0.63$, indicating that they do not represent independent directions in feature space but a single underlying physical dimension: the UV--to--optical/NIR spectral slope associated with recent star formation. In contrast, F277W--F444W shows only weak correlations with this u--* group (typically $|r| \lesssim 0.2$), confirming that it provides a largely independent constraint on the rest-frame optical/NIR color and hence on the age/metallicity of the dominant stellar population.

\subsection{Application to the COSMOS2025 sample}\label{sec:6.3}
Having validated the classifier and examined its feature behavior in the SAM, we next apply the trained \texttt{CatBoostClassifier} to the COSMOS2025 sample. To ensure strict consistency between the domains, we first construct the same set of 66 colors from the twelve common bands (Table~\ref{tab:tab1}) used for training, and feed these as inputs to the model. As for the SAM, the classifier returns both a hard morphology label (using the default probability threshold of 0.5) and a probabilistic classification $\text{P}(\mathrm{\text{Early type}})$ for every COSMOS2025 galaxy.

\begin{figure*}
	\centering
	\includegraphics[width=\linewidth]{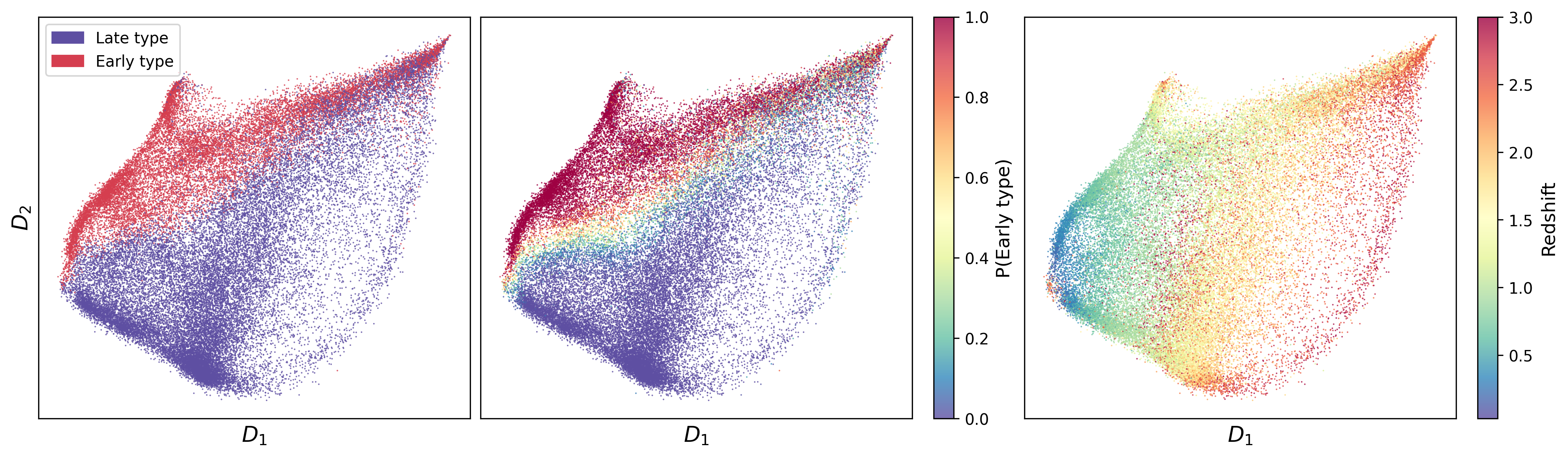}
	\caption{UMAP projection of the COSMOS2025 sample in color space after applying the trained \texttt{CatBoostClassifier}. Left: Hard morphology labels, showing galaxies classified as late-type (blue) and early-type (red). Middle: The same projection colored by the predicted probability of being early-type. Right: The projection colored by redshift.}
	\label{fig:fig9}
\end{figure*}

\begin{figure*}
	\centering
	\includegraphics[width=\linewidth]{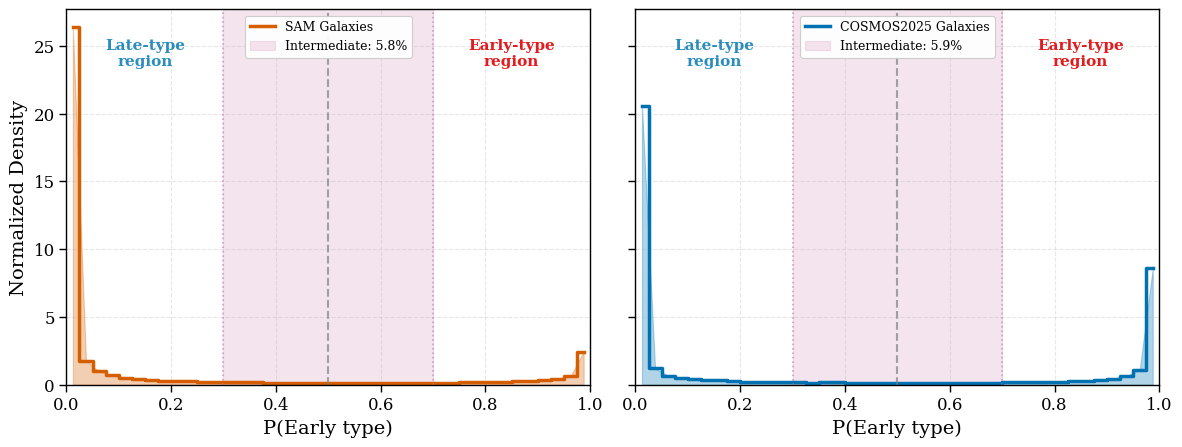}
	\caption{Distribution of early-type probabilities $\text{P}(\mathrm{\text{Early type}})$ for (left) the full SAM sample and (right) the COSMOS2025 sample.}
	\label{fig:prob_dist_comparison}
\end{figure*}

Applying the model to the 44{,}132-object COSMOS2025 sample yields 29{,}044 galaxies classified as late-type and 15{,}088 as early-type. Figure~\ref{fig:fig9} summarizes these results in the 2D UMAP space built from the COSMOS2025 colors: the left panel shows the hard early- and late-type assignments, the middle panel displays the corresponding early-type probabilities, and the right panel illustrates how these probabilities vary smoothly with redshift across the main color locus.

The probabilistic outputs allow us to quantitatively compare the structure of the COSMOS2025 color space with that of the SAM. Figure~\ref{fig:prob_dist_comparison} shows the distribution of $\text{P}(\mathrm{\text{Early type}})$ for the full COSMOS2025 sample alongside the SAM distribution from Section~\ref{sec:6.1}. The two distributions are strikingly similar: only 5.9\% of COSMOS2025 galaxies fall in the intermediate probability range $0.3 < \text{P}(\mathrm{\text{Early type}}) < 0.7$, nearly identical to the 5.8\% observed in the SAM. The remaining 94\% of galaxies in both domains receive high-confidence classifications ($\text{P}(\mathrm{\text{Early type}}) \leq 0.3$ or $\geq 0.7$).

This result has two implications. First, it suggests that broadband colors are an effective morphological discriminant in both simulated and observed populations: the majority of galaxies---including many with intermediate structural properties---have integrated colors that place them close to one photometric archetype. Second, the close agreement between the SAM and COSMOS2025 distributions supports the effectiveness of our simulation-based training and noise-injection approach, indicating that the classifier has learned a mapping from colors to morphology that transfers across the domain gap with minimal systematic offset.

\subsection{Validation with COSMOS2025 Bulge+Disk Decompositions}
To validate the performance of our simulation-trained classifier on real observational data, we compare its predictions against independent structural morphology measurements from the COSMOS2025 catalog. The catalog provides bulge-to-total flux ratios ($\mathrm{\text{B/T}}$) derived from two-dimensional S\'ersic profile fitting to the high-resolution JWST/NIRCam images \citep{shuntov2025cosmos2025}. These measurements offer a direct, observationally-derived estimate of galaxy structural morphology that is independent of the photometric colors used by our classifier.

We use the $\mathrm{\text{B/T}_{\text{F277W}}}$ measurement as our primary validation metric. This band selection is particularly appropriate as the F277W-F444W color emerged as the most important feature in our classifier (Figure~\ref{fig:fig8}), while maintaining high signal-to-noise detections. We adopt $\mathrm{\text{B/T}_{\text{F277W}}} \geq 0.4$ and $\mathrm{\text{B/T}_{\text{F277W}}} < 0.4$ to define early- and late-type galaxies, respectively.

\begin{figure*}
	\centering
	\includegraphics[width=\linewidth]{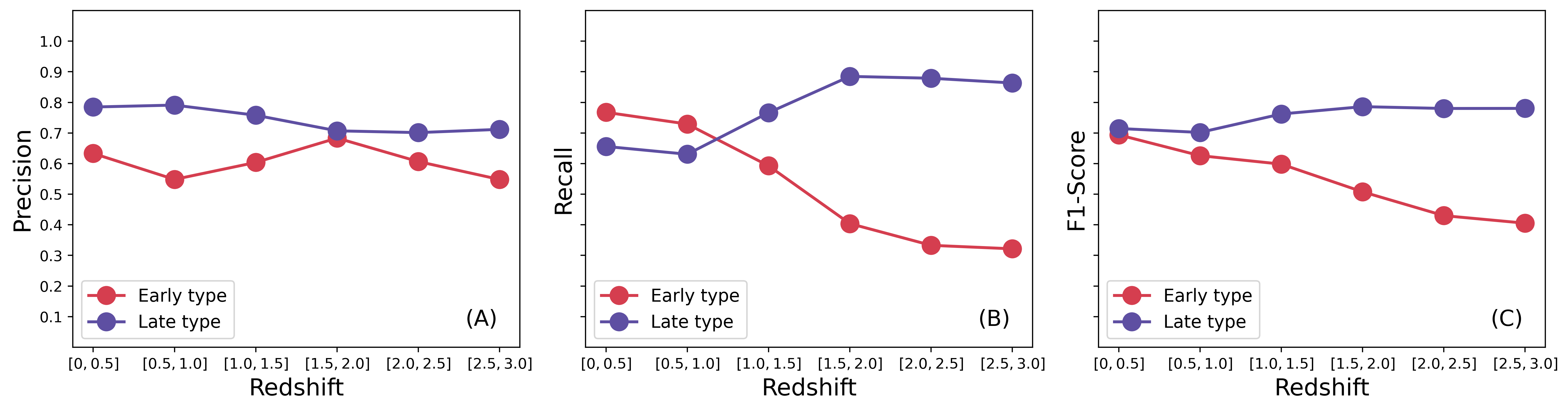}
	\caption{Redshift-dependent validation of the ML classifier against COSMOS2025 $\mathrm{\text{B/T}_{\text{F277W}}}$-based structural classifications.}
	\label{fig:fig88}
\end{figure*}

Using the $\mathrm{\text{B/T}_{\text{F277W}}}$-based early- and late-type split as an external benchmark, the COSMOS2025 classifications from our ML model achieve an overall accuracy of 70\% in reproducing the structural morphology labels. The early-type class is recovered with a precision (purity) of 54\% and a recall (completeness) of 59\%, while late-type galaxies are identified with a precision of 78\% and a recall of 74\%, indicating that the classifier performs particularly well for disk-dominated systems but still provides reasonably consistent predictions for bulge-dominated galaxies given the continuous nature of galaxy structure. These results are summarized in Table~\ref{tab:cosmos_bt_confusion}.

\begin{table}[t]
	\centering
	\begin{tabular}{c|ccc}
		\hline \hline
		Class & Precision [\%] & Recall [\%] & F1-Score [\%] \\
		\hline
		Late-type & 77.7 & 73.7 & 75.7 \\
		Early-type & 53.9 & 59.3 & 56.5 \\
		\hline
	\end{tabular}
	\caption{Performance of the ML classifier on the COSMOS2025 sample when using the $\mathrm{\text{B/T}_{\text{F277W}}}$ = 0.4 threshold as the reference structural classification.}
	\label{tab:cosmos_bt_confusion}
\end{table}

With a closer look at the redshift dependence, classifier performance reveals pronounced divergence across morphological classes. Figure~\ref{fig:fig88} presents the precision, recall, and F1-score as a function of redshift for both morphological classes, evaluated in six bins spanning $0 < \text{z} < 3$. For late-type galaxies, the classifier maintains consistently precision (71--79\%) across all epochs, while recall systematically improves from $\sim$66\% at $\text{z} < 0.5$ to $\sim$88\% at $\text{z} > 2.5$. This increasing completeness at higher redshifts likely reflects the growing dominance of disk-dominated, star-forming systems during the peak epoch of cosmic star formation, where the color-sSFR-structure relation becomes tighter and the classifier's photometric decision boundaries more effective.

In contrast, early-type galaxies exhibit markedly different redshift dependence. Precision remains relatively stable, but recall shows a systematic and substantial degradation from $\sim$77\% at $\text{z} < 0.5$ to only $\sim$33\% at $2.5 \leq \text{z} < 3.0$. This trend directly mirrors the behavior observed in the SAM testing set (Figure~\ref{fig:fig6}), confirming that the simulation-trained classifier has successfully transferred the underlying physical redshift dependence to the observational domain rather than introducing spurious domain-transfer artifacts.

Although the agreement between our ML classifications and the $\mathrm{\text{B/T}_{\text{F277W}}}$-based structural labels is at the $\sim$70\% level in terms of overall accuracy, the method is extremely fast: the full training and prediction steps require less than 5 minutes, while including photometric preprocessing and hyperparameter optimization the end-to-end pipeline completes in under 30 minutes on a standard workstation equipped with an 11th Gen Intel\textsuperscript{\textregistered} Core\texttrademark{} i7-1165G7 processor. This demonstrates that a simulation-trained classifier can provide reasonably accurate morphological information at a fraction of the computational cost of full bulge+disk decompositions.

\section{Discussion}\label{sec:7}
Our study presents a simulation-trained, color-based ML framework for classifying galaxies into early- and late-type morphological classes across a broad redshift range (\(0 < \text{z} < 3\)). By leveraging the physical realism of semi-analytic models and the photometric richness of the COSMOS2025 catalog, we have demonstrated that a relatively simple, interpretable classifier---trained on SAM-derived colors and applied to real multi-band photometry---can recover morphological information with reasonable accuracy and significant computational efficiency. 

\subsection{Performance in the SAM Domain}
Within the SAM, the classifier achieves near-saturated performance for late-type galaxies and robust, though slightly lower, recovery for early-type systems, on the secure testing set. This indicates that, in a controlled environment where the mapping between structure, star formation, and colors is internally consistent, broad-band colors alone contain enough information to separate bulge-dominated quiescent galaxies from disk-dominated star-forming galaxies with high reliability \citep[e.g., ][]{strateva2001color,bell2003optical,taylor2015galaxy}.

The mild decline in early-type recall with increasing redshift, contrasted with the essentially redshift-independent late-type performance, reflects the growing complexity of galaxy populations at early cosmic times. In particular, the SAM predicts a larger fraction of structurally intermediate or morphologically disturbed systems at $\text{z} \gtrsim 2$, for which the traditional early–late dichotomy becomes less sharp and color-based decision boundaries are intrinsically more ambiguous \citep[e.g., ][]{lotz2004new,conselice2014evolution}. The persistence of high early-type precision even in the highest redshift bins suggests that the classifier remains conservative in assigning early-type labels, favoring purity over completeness when the color distributions of the two classes overlap.

\subsection{Performance in the COSMOS2025 Domain}
The probabilistic classifications obtained for COSMOS2025 exhibit the same strong bimodality observed in the SAM domain (Figure~\ref{fig:prob_dist_comparison}). Galaxies in both simulated and observed populations cluster decisively around the early- and late-type photometric archetypes, with only a small fraction occupying the intermediate regime.

This close agreement between the SAM and COSMOS2025 probability distributions provides empirical support for our domain-transfer strategy. The photometric bimodality captured by the classifier is not an artifact of the simulation, but a genuine feature of real galaxy populations.

A key insight emerging from this probabilistic framework is the distinction between structural and photometric ambiguity. Although a substantial fraction of galaxies in the SAM are classified as structurally ambiguous based on B/T and sSFR criteria, the vast majority nonetheless receive decisive probability assignments. This reveals that integrated broadband colors provide a cleaner mapping onto the early–late dichotomy than structural cuts alone. Most galaxies with intermediate bulge fractions or mismatched morphology and star formation have colors that align them clearly with one photometric archetype.

The small fraction of galaxies with intermediate probabilities are therefore not methodological artifacts or classification failures. Rather, they represent the genuine photometric ``gray zone''---systems whose integrated colors reflect mixed stellar populations, ongoing quenching, or complex structural states. As such, they constitute a scientifically valuable sample for targeted follow-up studies of morphological transformation and the quenching process.

\subsection{Comparison with COSMOS2025 Bulge+Disk Morphologies}
When transferred to COSMOS2025, the simulation-trained classifier recovers the F277W-based bulge+disk classifications with an overall accuracy of roughly seven-tenths, yielding a substantially higher success rate for late-type than for early-type systems. This level of agreement is notable given that the two schemes rely on fundamentally different information---integrated colors versus two-dimensional light-profile decompositions \citep[e.g., ][]{peng2002detailed,simard2011catalog}---and that real galaxies exhibit structural and star-formation diversity that is necessarily simplified in the SAM.

The better performance for late-type systems likely reflects the relatively tight coupling between blue colors, high sSFR, and disk-dominated structure, whereas red colors can arise from a mixture of bulge-dominated quenched galaxies, dust-reddened star-forming disks, and composite systems with significant bulge and disk components \citep[e.g., ][]{muzzin2013evolution,ilbert2013mass}. Disagreements between the two classifications therefore do not necessarily represent outright failures, but instead highlight galaxies that are structurally intermediate or that deviate from the SAM’s assumed relation between morphology and star-formation history, and thus may be of particular interest for follow-up studies of morphological transformation \citep[e.g., ][]{martin2007uv,schawinski2014green}.

\subsection{Advantages and Use Cases}
A key strength of this framework is its computational efficiency: training and inference on tens of thousands of objects complete in minutes on a standard workstation (For a more detailed discussion, see Section 5.2.1 in \cite{asadi2025leveraging}), compared with the far greater resources required for full bulge+disk decompositions or deep CNN-based image classifiers. Combined with the interpretability of color-based decision boundaries and feature importance rankings, this makes the method well suited for constructing large, approximate morphological catalogs in upcoming wide-area surveys where detailed structural measurements will be infeasible for the full galaxy population.

The probabilistic outputs further enable flexible sample definitions tailored to specific science goals. For example, users interested in highly pure early-type samples can adopt a high probability threshold, while studies that prioritize completeness over purity can choose more inclusive cuts or work directly with the continuous early-type probability as a weight in statistical analyses.

\subsection{Limitations and Future Directions}
While our framework offers significant speed and interpretability advantages, it is subject to limitations inherent to both the method and the underlying simulation. First, the classifier relies on the physical recipes of the Santa Cruz SAM; any deviation between the simulated and real mapping of galaxy colors to morphology---arising from dust attenuation modeling or star-formation prescriptions---will introduce systematic biases. Second, by utilizing integrated colors alone, we discard spatial information such as asymmetry and concentration, which are valuable for identifying mergers or distinguishing between face-on disks and spheroids.

A critical physical limitation of this study is the restriction to $\text{z} < 3$. We deliberately adopted this cut because the traditional early- versus late-type dichotomy becomes increasingly ill-defined at higher redshifts. Beyond $\text{z} \sim 3$, galaxy assembly is dominated by irregular and clumpy structures where the classical Hubble sequence is not yet fully established \citep[e.g., ][]{elmegreen2007resolved,guo2015clumpy}. In this regime, apparent ``bulges" may often represent transient clumps formed via violent disk instabilities rather than merger-built spheroids, rendering standard B/T ratios ambiguous \citep[e.g., ][]{dekel2009cold,ceverino2010high}. Furthermore, at $\text{z} > 4$, many galaxies are still in the process of stabilizing their disks, making a simple structural binary insufficient to capture the diversity of the population.

However, the advent of JWST has fundamentally altered the landscape for high-redshift morphology, revealing rotationally supported disks and surprisingly mature structures at $\text{z} \sim 4-6$ that were previously inaccessible to HST \citep[e.g., ][]{ferreira2022panic,casey2023cosmos,robertson2023identification}. To extend our ML framework into this era of galaxy assembly, future work must evolve beyond purely photometric structural labels. A robust high-redshift classifier will require a hybrid morphological-kinematic approach: rather than relying solely on B/T, training labels should incorporate kinematic diagnostics available in the simulation---such as the rotation-to-dispersion ratio ($\text{v}/\sigma$)---to identify rotationally supported disks that may appear featureless or clumpy in imaging \citep[e.g., ][]{wisnioski2015kmos3d,amvrosiadis2025kinematics}. Additionally, the classification scheme should be expanded to include a specialized ``irregular or clumpy" class, allowing the model to explicitly identify the complex, transitioning systems that characterize the peak epochs of cosmic assembly.

\section{Conclusion}\label{sec:8}
We have presented a simulation-based ML framework to classify galaxies into early- and late-type morphologies out to $\text{z} \sim 3$, utilizing the COSMOS2025 catalog. By training a \texttt{CatBoostClassifier} on 66 photometric color features derived from the Santa Cruz SAM and injecting realistic observational noise, we bridged the gap between theoretical predictions and real observational data without relying on expensive visual or structural classification labels. Our main findings are as follows:

\begin{itemize}
	\item [--] In the simulated domain, our CatBoost classifier achieves excellent performance: late-type galaxies are recovered with 98\% recall and precision, while early-types show 88\% recall and 91\% precision. Early-type recall shows a mild redshift dependence, declining to $\sim$74\% at $\text{z} \sim 2.5-3.0$ while maintaining high purity.
	
	\item [--] Applied to the 44,132 galaxies in COSMOS2025, the model classifies 15,088 as early-type and 29,044 as late-type. Compared to structural B/T-based classifications, it achieves 70\% overall accuracy, with late-types identified at 78\% purity and 74\% completeness, and early-types at 54\% purity and 59\% completeness.
	
	\item [--] The classifier's probabilistic outputs reveal a strong bimodality in both simulated and observed galaxy populations: only $\sim$6\% of galaxies in SAM and COSMOS2025 receive intermediate probabilities ($0.3 < \text{P}(\mathrm{\text{Early type}}) < 0.7$), while the remaining 94\% are classified with high confidence. This demonstrates that broadband colors are a decisive morphological discriminant, and that most structurally ambiguous galaxies are photometrically unambiguous. The small intermediate population represents genuine transitional systems—prime targets for follow-up studies of galaxy transformation.
	
	\item [--] The classifier demonstrates physical interpretability, with the F277W-F444W color emerging as the most important feature---consistent with the rest-frame optical/NIR contrast between old and young stellar populations. The redshift-dependent performance in observations closely mirrors that in simulations, validating successful domain transfer.
	
	\item [--] The entire pipeline—from data preprocessing through training to prediction on tens of thousands of galaxies—completes in under 30 minutes on standard desktop hardware.
\end{itemize}

This work demonstrates that simulation-trained color-based classifiers provide a fast, interpretable, and reasonably accurate approach to morphological classification for large galaxy surveys. By leveraging physically motivated training data and focusing on photometric colors, we bridge the gap between idealized simulations and observational datasets, offering a scalable solution for next-generation surveys where detailed structural measurements remain computationally prohibitive for full samples.

\section*{Acknowledgments}
We thank the anonymous referee for their thorough and constructive comments, which have significantly strengthened the methodological rigor and clarity of this manuscript. The corresponding author gratefully acknowledges Prof. Hosein Haghi and Dr. Akram Hasani Zonoozi for their invaluable guidance and mentorship during the PhD studies that provided part of the scientific foundation for this work.

\section*{Data Availability}
The trained ML model and full classified COSMOS2025 sample (including early- and late-type galaxy classifications) resulting from this analysis are publicly available in the GitHub repository: \url{https://github.com/vahidoo7/COSMOS2025-Early-Late-Galaxy-Classifier}.

\software{Matplotlib \citep{barrett2005matplotlib}, Pandas \citep{mckinney2011pandas}, Scikit-learn \citep{pedregosa2011scikit}, Astropy \citep{robitaille2013astropy}, UMAP \citep{mcinnes2018umap}, Numpy \citep{harris2020array}, SciPy \citep{virtanen2020scipy}, Seaborn \citep{waskom2021seaborn}.}

\bibliography{ref}

@article{mcinnes2018umap,
  author={McInnes, Leland and Healy, John and Saul, Nathaniel and Großberger, Lukas},
  journal={\href{https://doi.org/10.48550/arXiv.1802.03426}{arXiv preprint arXiv:1802.03426}},
  year=2018}

@article{somerville2015physical,
  author={Somerville, Rachel S and Dav{\'e}, Romeel},
  journal = {\href{https://www.annualreviews.org/content/journals/10.1146/annurev-astro-082812-140951}{ARA\&A, 53, 51}},
  year = {2015}}

@article{asadi2025leveraging,
  author={Asadi, Vahid and Zonoozi, Akram Hasani and Haghi, Hosein and Abedini, Fatemeh and Kalantari, Atousa and Jafariyazani, Marziye and Chartab, Nima},
  journal={\href{https://iopscience.iop.org/article/10.3847/1538-4357/ade805}{ApJ, 989, 65}},
  year     = 2025}

@article{humphrey2023euclid,
  author={Humphrey, A and Bisigello, L and Cunha, PAC and Bolzonella, M and Fotopoulou, S and Caputi, K and Tortora, C and Zamorani, G and Papaderos, P and Vergani, D and others},
  journal ={\href{https://www.aanda.org/articles/aa/abs/2023/03/aa44307-22/aa44307-22.html}{A\&A, 671, A99}},
  year={2023}}

@article{weaver2022cosmos2020,
  author={Weaver, John R and Kauffmann, OB and Ilbert, Olivier and McCracken, Henry J and Moneti, Andrea and Toft, Sune and Brammer, Gabriel and Shuntov, Marko and Davidzon, Iary and Hsieh, Bau-Ching and others},
  journal ={\href{https://iopscience.iop.org/article/10.3847/1538-4365/ac3078/meta}{ApJS, 258, 11}},
  year={2022}}

@article{yung2022semi,
  author={Yung, LY Aaron and Somerville, Rachel S and Ferguson, Henry C and Finkelstein, Steven L and Gardner, Jonathan P and Dav{\'e}, Romeel and Bagley, Micaela B and Popping, Gerg{\"o} and Behroozi, Peter},
  journal ={\href{https://academic.oup.com/mnras/article-abstract/515/4/5416/6652501}{MNRAS, 515, 5416}},
  year={2022}}

@article{somerville2021mock,
  author={Somerville, Rachel S and Olsen, Charlotte and Yung, LY Aaron and Pacifici, Camilla and Ferguson, Henry C and Behroozi, Peter and Osborne, Shannon and Wechsler, Risa H and Pandya, Viraj and Faber, Sandra M and others},
  journal ={\href{https://academic.oup.com/mnras/article-abstract/502/4/4858/6123877}{MNRAS, 502, 4858}},
  year={2021}}

@article{oke1983secondary,
  author={Oke, JB and Gunn, JE},
  journal ={\href{https://adsabs.harvard.edu/full/1983ApJ...266..713O}{ApJ, 266, 713}},
year={1983}}

@article{klypin2011dark,
  author={Klypin, Anatoly A and Trujillo-Gomez, Sebastian and Primack, Joel},
  journal ={\href{https://iopscience.iop.org/article/10.1088/0004-637X/740/2/102/meta}{ApJ, 740, 102}},
  year={2011}}

@article{somerville2008semi,
  author={Somerville, Rachel S and Hopkins, Philip F and Cox, Thomas J and Robertson, Brant E and Hernquist, Lars},
  journal ={\href{https://academic.oup.com/mnras/article-abstract/391/2/481/1079250}{MNRAS, 391, 481}},
year={2008}}

@article{somerville2015star,
  author={Somerville, Rachel S and Popping, Gerg{\"o} and Trager, Scott C},
  journal ={\href{https://academic.oup.com/mnras/article-abstract/453/4/4337/2593703}{MNRAS, 453, 4337}},
  year={2015}}

@article{robertson2006fundamental,
  author={Robertson, Brant and Cox, Thomas J and Hernquist, Lars and Franx, Marijn and Hopkins, Philip F and Martini, Paul and Springel, Volker},
  journal ={\href{https://iopscience.iop.org/article/10.1086/500360/meta}{ApJ, 441, 21}},
  year={2006}}

@article{hopkins2009disks,
  author={Hopkins, Philip F and Cox, Thomas J and Younger, Joshua D and Hernquist, Lars},
  journal ={\href{https://iopscience.iop.org/article/10.1088/0004-637X/691/2/1168/meta}{ApJ, 691, 1168}},
  year={2009}}

@article{bondi1952spherically,
  author={Bondi, HJ1952MNRAS},
  journal ={\href{https://academic.oup.com/mnras/article-abstract/112/2/195/2601964}{MNRAS, 112, 195}},
  year={1952}}

@article{yung2019semi,
  author={Yung, LY Aaron and Somerville, Rachel S and Finkelstein, Steven L and Popping, Gerg{\"o} and Dav{\'e}, Romeel},
  journal ={\href{https://academic.oup.com/mnras/article-abstract/483/3/2983/5218517}{MNRAS, 483, 2983}},
  year={2019}}

@article{madau1995radiative,
  author={Madau, Piero},
  journal ={\href{https://adsabs.harvard.edu/full/record/seri/ApJ../0441/1995ApJ...441...18M.html}{ApJ, 441, 18}},
  year={1995}}

@article{scoville2007cosmic,
  author={Scoville, Nick and Aussel, H and Brusa, Marcella and Capak, Peter and Carollo, C Marcella and Elvis, M and Giavalisco, M and Guzzo, L and Hasinger, G and Impey, C and others},
  journal ={\href{https://iopscience.iop.org/article/10.1086/516585/meta}{ApJS, 172, 1}},
  year={2007}}

@article{laigle2016cosmos2015,
  author={Laigle, Clotilde and McCracken, Henry J and Ilbert, Olivier and Hsieh, Bau-Ching and Davidzon, Iary and Capak, Peter and Hasinger, G{\"u}nther and Silverman, John D and Pichon, Christophe and Coupon, Jean and others},
  journal ={\href{https://iopscience.iop.org/article/10.3847/0067-0049/224/2/24/meta}{ApJS, 224, 24}},
  year={2016}}

@article{mccracken2012ultravista,
  author={McCracken, HJ and Milvang-Jensen, B and Dunlop, J and Franx, M and Fynbo, JPU and Le F{\`e}vre, O and Holt, J and Caputi, KI and Goranova, Y and Buitrago, F and others},
  journal ={\href{https://www.aanda.org/articles/aa/abs/2012/08/aa19507-12/aa19507-12.html}{A\&A, 544, A156}},
  year={2012}}

@article{pacifici2016evolution,
  author={Pacifici, Camilla and Kassin, Susan A and Weiner, Benjamin J and Holden, Bradford and Gardner, Jonathan P and Faber, Sandra M and Ferguson, Henry C and Koo, David C and Primack, Joel R and Bell, Eric F and others},
  journal ={\href{https://iopscience.iop.org/article/10.3847/0004-637X/832/1/79/meta}{ApJ, 832, 79}},
  year={2016}}

@article{stekhoven2015missforest,
  author={Stekhoven, Daniel J},
  journal ={\href{https://ui.adsabs.harvard.edu/abs/2015ascl.soft05011S/abstract}{Astrophysics Source Code Library, ascl:1505}},
  year={2015}}

@article{prokhorenkova2018catboost,
  author={Prokhorenkova, Liudmila and Gusev, Gleb and Vorobev, Aleksandr and Dorogush, Anna Veronika and Gulin, Andrey},
  journal ={\href{https://proceedings.neurips.cc/paper/2018/hash/14491b756b3a51daac41c24863285549-Abstract.html}{NIPS, 31}},
  year={2018}}

@article{arnouts1999measuring,
  author={Arnouts, Stephane and Cristiani, Stefano and Moscardini, Lauro and Matarrese, Sabino and Lucchin, Francesco and Fontana, Adriano and Giallongo, Emanuele},
  journal ={\href{https://academic.oup.com/mnras/article-abstract/310/2/540/1048431}{MNRAS, 310, 540}},
  year={1999}}

@article{ilbert2006accurate,
  author={Ilbert, Olivier and Arnouts, S and Mccracken, Henry J and Bolzonella, M and Bertin, Emmanuel and Le F{\`e}vre, Olivier and Mellier, Yannick and Zamorani, G and Pello, R and Iovino, Angela and others},
  journal ={\href{https://www.aanda.org/articles/aa/abs/2006/39/aa5138-06/aa5138-06.html}{A\&A, 457, 841}},
  year={2006}}

@article{stehman1997selecting,
  author={Stehman, Stephen V},
  journal ={\href{https://www.sciencedirect.com/science/article/pii/S0034425797000837}{Remote sensing of Environment, 62, 77}},
  year={1997}}

@article{grogin2011candels,
  author={Grogin, Norman A and Kocevski, Dale D and Faber, SM and Ferguson, Henry C and Koekemoer, Anton M and Riess, Adam G and Acquaviva, Viviana and Alexander, David M and Almaini, Omar and Ashby, Matthew LN and others},
  journal ={\href{https://iopscience.iop.org/article/10.1088/0067-0049/197/2/35/meta}{ApJS, 197, 35}},
  year={2011}}

@article{koekemoer2011candels,
  author={Koekemoer, Anton M and Faber, SM and Ferguson, Henry C and Grogin, Norman A and Kocevski, Dale D and Koo, David C and Lai, Kamson and Lotz, Jennifer M and Lucas, Ray A and McGrath, Elizabeth J and others},
  journal ={\href{https://iopscience.iop.org/article/10.1088/0067-0049/197/2/36/meta}{ApJS, 197, 36}},
  year={2011}}

@article{gardner2006james,
  author={Gardner, Jonathan P and Mather, John C and Clampin, Mark and Doyon, Rene and Greenhouse, Matthew A and Hammel, Heidi B and Hutchings, John B and Jakobsen, Peter and Lilly, Simon J and Long, Knox S and others},
  journal ={\href{https://link.springer.com/article/10.1007/s11214-006-8315-7}{Space Science Reviews, 123, 485}},
  year={2006}}

@article{pedregosa2011scikit,
  author={Pedregosa, Fabian and Varoquaux, Ga{\"e}l and Gramfort, Alexandre and Michel, Vincent and Thirion, Bertrand and Grisel, Olivier and Blondel, Mathieu and Prettenhofer, Peter and Weiss, Ron and Dubourg, Vincent and others},
  journal={\href{http://www.jmlr.org/papers/volume12/pedregosa11a/pedregosa11a.pdf?source=post_page} {JMLR, 12, 2825}},
  year  = 2011}

@article{Breiman2001,
  author    = {Breiman, Leo},
  journal   = {\href{https://doi.org/10.1023/A:1010933404324}{Machine Learning, 45, 5}},
  year     = 2001}

@article{boulet2024catalogue,
  author={Boulet, Thibault},
  journal   = {\href{https://www.aanda.org/articles/aa/abs/2024/05/aa48031-23/aa48031-23.html}{A\&A, 685, A66}},
  year     = 2024}

@article{zeraatgari2024exploring,
  author={Zeraatgari, FZ and Hafezianzadeh, F and Zhang, Y-X and Mosallanezhad, A and Zhang, J-Y},
  journal   = {\href{https://www.aanda.org/articles/aa/abs/2024/08/aa48714-23/aa48714-23.html}{A\&A, 688, A33}},
  year     = 2024}

@article{coronado2022classification,
  author={Coronado-Bl{\'a}zquez, Javier},
  journal   = {\href{https://academic.oup.com/mnras/article-abstract/515/2/1807/6646522}{MNRAS, 515, 1807}},
  year     = 2022}

@article{coronado2023redshift,
  author={Coronado-Bl{\'a}zquez, Javier},
  journal   = {\href{https://academic.oup.com/mnras/article-abstract/521/3/4156/7080161}{MNRAS, 521, 4156}},
  year     = 2023}

@article{cunha2022photometric,
  author={Cunha, PAC and Humphrey, A},
  journal   = {\href{https://www.aanda.org/articles/aa/abs/2022/10/aa43135-22/aa43135-22.html}{A\&A, 666, A87}},
  year     = 2022}

@article{hughes2022quasar,
  author={Hughes, Arvind CN and Bailer-Jones, Coryn AL and Jamal, Sara},
  journal   = {\href{https://www.aanda.org/articles/aa/abs/2022/12/aa44859-22/aa44859-22.html}{A\&A, 668, A99}},
  year     = 2022}

@article{li2025application,
  author={Li, Jie and Lv, Hongkui and Liu, Yang and Huang, Jiajun and Wang, Yu and Lin, Wenbin},
  journal   = {\href{https://iopscience.iop.org/article/10.3847/1538-4365/ad9581/meta}{ApJS, 276, 24}},
  year     = 2025}

@article{mckinney2011pandas,
  author={McKinney, Wes and others},
  journal   = {\href{https://www.academia.edu/download/117768488/pyhpc2011_submission_9.pdf}{PyHPC, 14, 1}},
  year     = 2011}

@article{harris2020array,
  author={Harris, Charles R and Millman, K Jarrod and Van Der Walt, St{\'e}fan J and Gommers, Ralf and Virtanen, Pauli and Cournapeau, David and Wieser, Eric and Taylor, Julian and Berg, Sebastian and Smith, Nathaniel J and others},
  journal   = {\href{https://www.nature.com/articles/s41586-020-2649-2}{nature, 585, 357}},
  year     = 2020}

@article{robitaille2013astropy,
  author={Robitaille, Thomas P and Tollerud, Erik J and Greenfield, Perry and Droettboom, Michael and Bray, Erik and Aldcroft, Tom and Davis, Matt and Ginsburg, Adam and Price-Whelan, Adrian M and Kerzendorf, Wolfgang E and others},
  journal   = {\href{https://www.aanda.org/articles/aa/abs/2013/10/aa22068-13/aa22068-13.html}{A\&A, 558, A33}},
  year     = 2013}

@article{waskom2021seaborn,
  author={Waskom, Michael L},
  journal   = {\href{https://joss.theoj.org/papers/10.21105/joss.03021.pdf}{JOSS, 6, 3021}},
  year     = 2021}

@article{barrett2005matplotlib,
  author={Barrett, Paul and Hunter, John and Miller, J Todd and Hsu, J-C and Greenfield, Perry},
  journal   = {\href{https://adsabs.harvard.edu/full/2005ASPC..347...91B}{ADASS XIV, 347, 91}},
  year     = 2005}

@article{virtanen2020scipy,
  author={Virtanen, Pauli and Gommers, Ralf and Oliphant, Travis E and Haberland, Matt and Reddy, Tyler and Cournapeau, David and Burovski, Evgeni and Peterson, Pearu and Weckesser, Warren and Bright, Jonathan and others},
  journal   = {\href{https://www.nature.com/articles/s41592-019-0686-2}{Nature methods, 17, 261}},
  year     = 2020}

@article{Asadi_2025,
  author={Asadi, Vahid and Chartab, Nima and Zonoozi, Akram Hasani and Haghi, Hosein and Gozaliasl, Ghassem and Haghjoo, Aryana and Mobasher, Bahram},
  journal   = {\href{https://iopscience.iop.org/article/10.3847/1538-4357/ae0a2c}{ApJ, 993, 123}},
  year     = 2025}

@book{geron2019hands,
  title={\href{https://www.oreilly.com/library/view/hands-on-machine-learning/9781098125967/}{Hands-On Machine Learning with Scikit-Learn, Keras, and TensorFlow: Concepts, Tools, and Techniques to Build Intelligent Systems}},
  author={G{\'e}ron, Aur{\'e}lien},
  year={2022},
  publisher={O'Reilly Media},
  edition={3nd}}

@article{shuntov2025cosmos2025,
  author={Shuntov, Marko and Akins, Hollis B and Paquereau, Louise and Casey, Caitlin M and Ilbert, Olivier and Arango-Toro, Rafael C and McCracken, Henry Joy and Franco, Maximilien and Harish, Santosh and Kartaltepe, Jeyhan S and others},
  journal   = {\href{https://arxiv.org/abs/2506.03243}{arXiv preprint arXiv:2506.03243}},
  year     = 2025}

@article{boquien2019cigale,
  author={Boquien, M and Burgarella, D and Roehlly, Y and Buat, V and Ciesla, L and Corre, D and Inoue, AK and Salas, H},
  journal   = {\href{https://iopscience.iop.org/article/10.3847/1538-4357/ae0a2c}{A\$A, 622, A103}},
  year     = 2019}

@article{allen2006millennium,
  author={Allen, Paul D and Driver, Simon P and Graham, Alister W and Cameron, Ewan and Liske, Jochen and De Propris, Roberto},
  journal={\href{https://academic.oup.com/mnras/article-abstract/371/1/2/979080}{MNRAS, 371, 2}},
  year= 2006}

@article{de1996near,
  author={de Jong, Roelof S},
  journal={\href{https://inspirehep.net/legacy/arxiv/astro-ph/9601005}{A\&A, 313, 45}},
  year= 1996}

@article{hubble1926extragalactic,
  author={Hubble, Edwin P},
  journal   = {\href{https://adsabs.harvard.edu/full/1926ApJ....64..321H7}{ApJ, 64, 321}},
  year     = 1926}

@article{baldry2004quantifying,
  author={Baldry, Ivan K and Glazebrook, Karl and Brinkmann, Jon and Ivezi{\'c}, {\v{Z}}eljko and Lupton, Robert H and Nichol, Robert C and Szalay, Alexander S},
  journal   = {\href{https://iopscience.iop.org/article/10.1086/380092/meta}{ApJ, 600, 681}},
  year     = 2004}

@article{conselice2014evolution,
  author={Conselice, Christopher J},
  journal   = {\href{https://www.annualreviews.org/content/journals/10.1146/annurev-astro-081913-040037}{ARA, 52, 291}},
  year     = 2014}

@article{mortlock2013redshift,
  author={Mortlock, Alice and Conselice, Christopher J and Hartley, William G and Ownsworth, Jamie R and Lani, Caterina and Bluck, Asa FL and Almaini, Omar and Duncan, Kenneth and Wel, Arjen van der and Koekemoer, Anton M and others},
  journal={\href{https://academic.oup.com/mnras/article-abstract/433/2/1185/1747723}{MNRAS, 433, 1185}},
  year= 2013}

@article{huertas2016mass,
  author={Huertas-Company, M and Bernardi, M and P{\'e}rez-Gonz{\'a}lez, PG and Ashby, MLN and Barro, G and Conselice, C and Daddi, E and Dekel, A and Dimauro, P and Faber, SM and others},
  journal={\href{https://academic.oup.com/mnras/article-abstract/462/4/4495/2589496}{MNRAS, 462, 4495}},
  year= 2016}

@article{lintott2008galaxy,
  author={Lintott, Chris J and Schawinski, Kevin and Slosar, An{\v{z}}e and Land, Kate and Bamford, Steven and Thomas, Daniel and Raddick, M Jordan and Nichol, Robert C and Szalay, Alex and Andreescu, Dan and others},
  journal={\href{https://academic.oup.com/mnras/article-abstract/389/3/1179/1017183}{MNRAS, 389, 1179}},
  year= 2008}

@article{kartaltepe2015candels,
  author={Kartaltepe, Jeyhan S and Mozena, Mark and Kocevski, Dale and McIntosh, Daniel H and Lotz, Jennifer and Bell, Eric F and Faber, Sandy and Ferguson, Harry and Koo, David and Bassett, Robert and others},
  journal   = {\href{https://iopscience.iop.org/article/10.1088/0067-0049/221/1/11/meta}{ApJS, 221, 11}},
  year     = 2015}

@article{peng2002detailed,
  author={Peng, Chien Y and Ho, Luis C and Impey, Chris D and Rix, Hans-Walter},
  journal   = {\href{https://iopscience.iop.org/article/10.1086/340952/meta}{ApJ, 124, 266}},
  year     = 2002}

@article{simard2011catalog,
  author={Simard, Luc and Mendel, J Trevor and Patton, David R and Ellison, Sara L and McConnachie, Alan W},
  journal   = {\href{https://iopscience.iop.org/article/10.1088/0067-0049/196/1/11/meta}{ApJS, 196, 11}},
  year     = 2011}

@article{vika2013megamorph,
  author={Vika, Marina and Bamford, Steven P and Haeussler, Boris and Rojas, Alex L and Borch, Andrea and Nichol, Robert C},
  journal={\href{https://academic.oup.com/mnras/article-abstract/435/1/623/1117369}{MNRAS, 435, 623}},
  year= 2013}

@article{dimauro2018catalog,
  author={Dimauro, Paola and Huertas-Company, Marc and Daddi, Emanuele and P{\'e}rez-Gonz{\'a}lez, Pablo G and Bernardi, Mariangela and Barro, Guillermo and Buitrago, Fernando and Caro, Fernando and Cattaneo, Andrea and Dominguez-S{\'a}nchez, Helena and others},
  journal={\href{https://academic.oup.com/mnras/article-abstract/478/4/5410/5004864}{MNRAS, 478, 5410}},
  year= 2018}

@article{nedkova2024bulge+,
  author={Nedkova, Kalina V and H{\"a}u{\ss}ler, Boris and Marchesini, Danilo and Brammer, Gabriel B and Feinstein, Adina D and Johnston, Evelyn J and Kartaltepe, Jeyhan S and Koekemoer, Anton M and Martis, Nicholas S and Muzzin, Adam and others},
  journal={\href{https://academic.oup.com/mnras/article-abstract/478/4/5410/5004864}{MNRAS, 532, 3747}},
  year= 2024}

@article{simmons2016galaxy,
  author={Simmons, Brooke D and Lintott, Chris and Willett, Kyle W and Masters, Karen L and Kartaltepe, Jeyhan S and H{\"a}u{\ss}ler, Boris and Kaviraj, Sugata and Krawczyk, Coleman and Kruk, SJ and McIntosh, Daniel H and others},
  journal={\href{https://academic.oup.com/mnras/article/464/4/4420/2417365}{MNRAS, 464, 4420}},
  year= 2016}

@article{willett2016galaxy,
  author={Willett, Kyle W and Galloway, Melanie A and Bamford, Steven P and Lintott, Chris J and Masters, Karen L and Scarlata, Claudia and Simmons, Brooke D and Beck, Melanie and Cardamone, Carolin N and Cheung, Edmond and others},
  journal={\href{https://academic.oup.com/mnras/article-abstract/464/4/4176/2527878}{MNRAS, 464, 4176}},
  year= 2016}

@article{cuillandre2025euclid,
  author={Cuillandre, J-C and Bertin, E and Bolzonella, M and Bouy, H and Gwyn, S and Isani, S and Kluge, M and Lai, O and Lan{\c{c}}on, A and Lang, DA and others},
  journal={\href{https://www.aanda.org/articles/aa/abs/2025/05/aa50803-24/aa50803-24.html}{A\&A, 697, A6}},
  year= 2025}

@article{wang2022high,
  author={Wang, Yun and Zhai, Zhongxu and Alavi, Anahita and Massara, Elena and Pisani, Alice and Benson, Andrew and Hirata, Christopher M and Samushia, Lado and Weinberg, David H and Colbert, James and others},
  journal={\href{https://iopscience.iop.org/article/10.3847/1538-4357/ac4973/meta}{ApJ, 928, 1}},
  year= 2022}

@article{cheng2021galaxy,
  author={Cheng, Ting-Yun and Conselice, Christopher J and Arag{\'o}n-Salamanca, Alfonso and Aguena, Michel and Allam, Sahar and Andrade-Oliveira, F and Annis, J and Bluck, AFL and Brooks, D and Burke, David L and others},
  journal={\href{https://academic.oup.com/mnras/article-abstract/507/3/4425/6327560}{MNRAS, 507, 4425}},
  year= 2021}

@article{Cao2024galaxy,
  author={Cao, Jie and Xu, Tingting and Deng, Yuhe and Deng, Linhua and Yang, Mingcun and Liu, Zhijing and Zhou, Weihong},
  journal={\href{https://www.aanda.org/articles/aa/abs/2024/03/aa48544-23/aa48544-23.html}{A\&A, 683, A42}},
  year= 2024}

@article{dominguez2018improving,
  author={Dom{\'\i}nguez S{\'a}nchez, H and Huertas-Company, M and Bernardi, M and Tuccillo, D and Fischer, JL},
  journal={\href{https://academic.oup.com/mnras/article-abstract/476/3/3661/4848300}{MNRAS, 476, 3661}},
  year= 2018}

@article{bhambra2022explaining,
  author={Bhambra, Prabh and Joachimi, Benjamin and Lahav, Ofer},
  journal={\href{https://academic.oup.com/mnras/article-abstract/511/4/5032/6529251}{MNRAS, 511, 5032}},
  year= 2022}

@article{fang2023automatic,
  author={Fang, GuanWen and Ba, Shuo and Gu, Yizhou and Lin, Zesen and Hou, Yuejie and Qin, Chenxin and Zhou, Chichun and Xu, Jun and Dai, Yao and Song, Jie and others},
  journal={\href{https://iopscience.iop.org/article/10.3847/1538-3881/aca1a6/meta}{ApJ, 165, 35}},
  year= 2023}

@article{iyer2024galaxy,
  author={Iyer, KG and Angeloudi, E and Bagley, MB and Finkelstein, SL and Kartaltepe, J and McGrath, EJ and Sarmiento, R and Vega-Ferrero, J and Haro, P Arrabal and Behroozi, P and others},
  journal={\href{https://www.aanda.org/articles/aa/abs/2025/05/aa50803-24/aa50803-24.html}{A\&A, 685, A48}},
  year= 2024}

@article{barchi2020machine,
  author={Barchi, Paulo H and de Carvalho, RR and Rosa, Reinaldo R and Sautter, RA and Soares-Santos, Marcelle and Marques, Bruno AD and Clua, Esteban and Gon{\c{c}}alves, TS and de S{\'a}-Freitas, C and Moura, TC},
  journal={\href{https://www.sciencedirect.com/science/article/pii/S2213133719300757}{Astronomy and Computing, 30, 100334}},
  year= 2020}

@article{csahin2025unlocking,
  author={{\c{S}}AHiN, Emrullah and Arslan, Naciye Nur and {\"O}zdemir, Durmu{\c{s}}},
  journal={\href{https://link.springer.com/article/10.1007/s00521-024-10437-2}{Neural Computing and Applications, 37, 859}},
  year= 2025}

@article{guo2015clumpy,
  author={Guo, Yicheng and Ferguson, Henry C and Bell, Eric F and Koo, David C and Conselice, Christopher J and Giavalisco, Mauro and Kassin, Susan and Lu, Yu and Lucas, Ray and Mandelker, Nir and others},
  journal={\href{https://iopscience.iop.org/article/10.1088/0004-637X/800/1/39/meta}{ApJ, 800, 39}},
  year= 2015}

@article{bell2003optical,
  author={Bell, Eric F and McIntosh, Daniel H and Katz, Neal and Weinberg, Martin D},
  journal   = {\href{https://iopscience.iop.org/article/10.1086/378847/meta}{ApJS, 149, 289}},
  year     = 2003}

@article{poggianti1997indicators,
  author={Poggianti, BM and Barbaro, G},
  journal={\href{https://research.rug.nl/en/publications/indicators-of-star-formation-4000-angstrom-break-and-balmer-lines}{A\&A, 325, 1025}},
  year= 1997}

@article{kauffmann2003stellar,
  author={Kauffmann, Guinevere and Heckman, Timothy M and White, Simon DM and Charlot, St{\'e}phane and Tremonti, Christy and Brinchmann, Jarle and Bruzual, Gustavo and Peng, Eric W and Seibert, Mark and Bernardi, Mariangela and others},
  journal={\href{https://academic.oup.com/mnras/article-abstract/341/1/33/999309}{MNRAS, 341, 33}},
  year= 2003}

@article{strateva2001color,
  author={Strateva, Iskra and Ivezi{\'c}, {\v{Z}}eljko and Knapp, Gillian R and Narayanan, Vijay K and Strauss, Michael A and Gunn, James E and Lupton, Robert H and Schlegel, David and Bahcall, Neta A and Brinkmann, Jon and others},
  journal={\href{https://iopscience.iop.org/article/10.1086/323301/meta}{ApJ, 122, 1861}},
  year= 2001}

@article{taylor2015galaxy,
  author={Taylor, Edward N and Hopkins, Andrew M and Baldry, Ivan K and Bland-Hawthorn, Joss and Brown, Michael JI and Colless, Matthew and Driver, Simon and Norberg, Peder and Robotham, Aaron SG and Alpaslan, Mehmet and others},
  journal={\href{https://academic.oup.com/mnras/article-abstract/446/2/2144/2891846}{MNRAS, 446, 2144}},
  year= 2015}

@article{lotz2004new,
  author={Lotz, Jennifer M and Primack, Joel and Madau, Piero},
  journal={\href{https://iopscience.iop.org/article/10.1086/421849/meta}{ApJ, 128, 163}},
  year= 2004}

@article{muzzin2013evolution,
  author={Muzzin, Adam and Marchesini, Danilo and Stefanon, Mauro and Franx, Marijn and McCracken, Henry J and Milvang-Jensen, Bo and Dunlop, James S and Fynbo, JPU and Brammer, Gabriel and Labb{\'e}, Ivo and others},
  journal={\href{https://iopscience.iop.org/article/10.1088/0004-637X/777/1/18/meta}{ApJ, 777, 18}},
  year= 2013}

@article{ilbert2013mass,
  author={Ilbert, Olivier and McCracken, Henry J and Le F{\`e}vre, Olivier and Capak, Peter and Dunlop, James and Karim, Alexander and Renzini, MA and Caputi, Karina and Boissier, Samuel and Arnouts, St{\'e}phane and others},
  journal={\href{https://www.aanda.org/articles/aa/abs/2013/08/aa21100-13/aa21100-13.html}{A\&A, 556, A55}},
  year= 2013}

@article{martin2007uv,
  author={Martin, D Christopher and Wyder, Ted K and Schiminovich, David and Barlow, Tom A and Forster, Karl and Friedman, Peter G and Morrissey, Patrick and Neff, Susan G and Seibert, Mark and Small, Todd and others},
  journal   = {\href{https://iopscience.iop.org/article/10.1086/516639/meta}{ApJS, 173, 342}},
  year     = 2007}

@article{schawinski2014green,
  author={Schawinski, Kevin and Urry, C Megan and Simmons, Brooke D and Fortson, Lucy and Kaviraj, Sugata and Keel, William C and Lintott, Chris J and Masters, Karen L and Nichol, Robert C and Sarzi, Marc and others},
  journal={\href{https://academic.oup.com/mnras/article-abstract/440/1/889/1749989}{MNRAS, 440, 889}},
  year= 2014}

@article{elmegreen2007resolved,
  author={Elmegreen, Debra Meloy and Elmegreen, Bruce G and Ravindranath, Swara and Coe, Daniel A},
  journal={\href{https://iopscience.iop.org/article/10.1086/511667/meta}{ApJ, 658, 763}},
  year= 2007}

@article{dekel2009cold,
  author={Dekel, A and Birnboim, Y and Engel, G and Freundlich, J and Goerdt, T and Mumcuoglu, M and Neistein, E and Pichon, C and Teyssier, R and Zinger, E},
  journal={\href{https://www.nature.com/articles/nature07648}{Nature, 457, 451}},
  year= 2009}

@article{ceverino2010high,
  author={Ceverino, Daniel and Dekel, Avishai and Bournaud, Frederic},
  journal={\href{https://academic.oup.com/mnras/article-abstract/404/4/2151/1089046}{MNRAS, 404, 2151}},
  year= 2010}

@article{ferreira2022panic,
  author={Ferreira, Leonardo and Adams, Nathan and Conselice, Christopher J and Sazonova, Elizaveta and Austin, Duncan and Caruana, Joseph and Ferrari, Fabricio and Verma, Aprajita and Trussler, James and Broadhurst, Tom and others},
  journal={\href{https://iopscience.iop.org/article/10.3847/2041-8213/ac947c/meta}{ApJ, 938, L2}},
  year= 2022}

@article{casey2023cosmos,
  author={Casey, Caitlin M and Kartaltepe, Jeyhan S and Drakos, Nicole E and Franco, Maximilien and Harish, Santosh and Paquereau, Louise and Ilbert, Olivier and Rose, Caitlin and Cox, Isabella G and Nightingale, James W and others},
  journal={\href{https://iopscience.iop.org/article/10.3847/1538-4357/acc2bc/meta}{ApJ, 954, 31}},
  year= 2023}

@article{robertson2023identification,
  author={Robertson, Brant E and Tacchella, S and Johnson, BD and Hainline, K and Whitler, L and Eisenstein, DJ and Endsley, R and Rieke, M and Stark, DP and Alberts, S and others},
  journal={\href{https://www.nature.com/articles/s41550-023-01921-1}{Nature, 7, 611}},
  year= 2023}

@article{wisnioski2015kmos3d,
  author={Wisnioski, E and Schreiber, NM F{\"o}rster and Wuyts, S and Wuyts, E and Bandara, K and Wilman, D and Genzel, R and Bender, R and Davies, R and Fossati, M and others},
  journal={\href{https://iopscience.iop.org/article/10.1088/0004-637X/799/2/209/meta}{ApJ, 799, 209}},
  year= 2015}

@article{amvrosiadis2025kinematics,
  author={Amvrosiadis, A and Wardlow, J\_L and Birkin, J\_E and Smail, I and Swinbank, A\_M and Nightingale, J and Bertoldi, F and Brandt, W\_N and Casey, C\_M and Chapman, S\_C and others},
  journal={\href{https://academic.oup.com/mnras/article-abstract/536/4/3757/7924248}{MNRAS, 536, 3757}},
  year= 2025}

@article{asadi_mass,
  author={Asadi, Vahid and Zonoozi, Akram Hasani and Haghi, Hosein},
  journal={\href{https://doi.org/10.3847/1538-4357/ae3144}{ApJ, 998, 2}},
  year= 2025}

@article{luo2025galaxy,
  author={Luo, Zhijian and Chen, Jianzhen and Chen, Zhu and Zhang, Shaohua and Fu, Liping and Xiao, Hubing and Shu, Chenggang},
  journal ={\href{https://iopscience.iop.org/article/10.3847/1538-4365/addb4c}{ApJS, 279, 17}},
  year={2025}}

@article{pandya2025sidda,
  author={Pandya, Sneh and Patel, Purvik and Nord, Brian D and Walmsley, Mike and {\'C}iprijanovi{\'c}, Aleksandra},
  journal ={\href{https://iopscience.iop.org/article/10.1088/2632-2153/adf701}{MLST, 6, 035032}},
  year={2025}}

@article{luo2024imputation,
  author={Luo, Zhijian and Tang, Zhirui and Chen, Zhu and Fu, Liping and Du, Wei and Zhang, Shaohua and Gong, Yan and Shu, Chenggang and Lu, Junhao and Li, Yicheng and others},
  journal ={\href{https://academic.oup.com/mnras/article/531/3/3539/7688471}{MNRAS, 531, 3539}},
  year={2024}}

@article{ren2020using,
  author={Ren, Bin and Pueyo, Laurent and Chen, Christine and Choquet, Elodie and Debes, John H and Duch{\^e}ne, Gaspard and M{\'e}nard, Fran{\c{c}}ois and Perrin, Marshall D},
  journal ={\href{https://iopscience.iop.org/article/10.3847/1538-4357/ab7024/meta}{ApJ, 892, 74}},
  year={2020}}

@article{keerin2022estimation,
  author={Keerin, Phimmarin and Boongoen, Tossapon},
  journal ={\href{https://www.sciencedirect.com/science/article/pii/S0306457322000127}{IP\&M, 59, 102881}},
  year={2022}}

@article{luo2024photometric,
  author={Luo, Zhijian and Li, Yicheng and Lu, Junhao and Chen, Zhu and Fu, Liping and Zhang, Shaohua and Xiao, Hubing and Du, Wei and Gong, Yan and Shu, Chenggang and others},
  journal ={\href{https://academic.oup.com/mnras/article/535/2/1844/7845879}{MNRAS, 535, 1844}},
  year={2024}}

\end{document}